\documentclass{article}
\newcommand{\be}{\begin{eqnarray}}
\newcommand{\ee}{\end{eqnarray}}
\newcommand{\nn}{\nonumber \\ \nonumber \\}
\newcommand{\nl}{\\  \nonumber \\}

\renewcommand{\vec}[1]{\mbox{\boldmath$#1$}}
\renewcommand{\d}{\mbox{\rm d}}

\begin{document}

\pagenumbering{arabic}

\title{Helicity Precession of Spin-1/2 Particles \\
in Weak Inertial and Gravitational Fields}
\author{Dinesh Singh$^{a}$\footnote{e-mail: singhd@uregina.ca},
Nader Mobed$^{a}$\footnote{e-mail: nader.mobed@uregina.ca}, Giorgio
Papini$^{a,b}$\footnote{e-mail: papini@uregina.ca} \\
$^a$Department of Physics, University of Regina,
Regina, Saskatchewan, S4S 0A2, Canada \\
$^b$International Institute for Advanced Scientific
Studies, 89019 Vietri sul Mare (SA), Italy}
\date{\today}

\maketitle

\begin{abstract}
We calculate the helicity and chirality effects experienced by a
spin-1/2 particle subjected to classical electromagnetic and
gravitational fields.  The helicity evolution is then determined
in the non-relativistic, relativistic, and ultra-relativistic
regimes. We find that inertia-gravitation can distinguish between
helicity and chirality. Helicity is not conserved, in general,
even when the particles are massless. In this case, however, the
inertial fields can hardly be applied to the fermions.

\end{abstract}

\setcounter{section}{0}
\setcounter{equation}{0}

\section{Introduction}

Over the past forty years, experimental connections between
inertia-gravitation and quantum mechanics have been established in
a limited number of instances. They confirm that inertia and
Newtonian gravitation affect particle wave-functions in ways that
are consistent with covariant generalizations of known wave
equations. Typical examples are represented by the Schr\"{o}dinger
and Klein-Gordon equations that have been successfully used to
describe the behaviour of superconducting electrons
\cite{DW,Papini1} and neutrons \cite{CaiPap} in
inertial-gravitational fields in a quasi-classical regime. Though
the lengths scales involved, of order $10^{-3}$ {\it cm} for
superelectrons \cite{Hild} and $10^{-13}$ {\it cm} for neutrons
\cite{COW}, are far from comparable to Planck's length,
which is thought to mark the onset of quantum gravity, the results
extend the validity of certain aspects of relativistic
inertia-gravitation by about thirty orders of magnitude.

Quantum particles are sensitive probes of inertia and, ultimately,
gravity \cite{GP}. The unavoidable presence of inertial effects in
precise tests of fundamental theories requires an attentive study
of all aspects of inertia.
As the length scales decrease, fundamental quantum properties of
particles, like spin and discrete symmetries, come into play
\cite{pap2}. This is the case for spin-1/2 particles, where they can be
used in a variety of experimental situations and energy ranges,
while still retaining a non-classical behaviour.

Particle accelerators and storage rings are very apt tools for
studying inertia, rotational inertia in particular. The forerunner
of these studies is the work by Bell and Leinaas \cite{Bell}.
They were able to calculate the effect of the coupling of spin to the
quantum contributions $\delta\omega$ to the angular velocity. This
is a kind of quantum Mashhoon effect that is likely responsible
for a residual electron depolarization in storage rings. Issues of
interest concern the behaviour of helicity \cite{Cai4} and
chirality in the presence of inertia. If $\omega$ is classical,
then spin has a precession frequency that equals the orbital
frequency for fermions with $g=2$. But when $\kappa \equiv (g-2)/2
\neq 0$, the spin vector undergoes an additional precession of
frequency $\kappa eB/m$ that is measured with extreme accuracy in
muon $g-2$ experiments \cite{BNL}. Persistent residual discrepancies between
standard model calculations and experiment have then been
interpreted by one of us \cite{pap2} as possible violations of the
discrete symmetries in rotational inertia. These would arise if
the gyro-gravitational ratio of the muon differs from one.
There is a similarity here with the electromagnetic case, where
$g=2$ is required by the Dirac equation, but not by quantum
electrodynamics. Some unresolved problems exist
regarding the helicity of massless particles due to rotational inertia,
which in Minkowski
space-time is a conserved quantity. This property may not hold
true in the presence of inertial-gravitational fields, as suggested
by Mashhoon for photons \cite{Mashhoon1,Mashhoon2}. Cai and Papini
\cite{Cai1,Papini2} found that spin-rotation coupling induces
oscillations between massive left-handed and right-handed
neutrinos (with non-vanishing magnetic moment), and that these
oscillations persist in the limit of vanishing mass. Mergulh\~{a}o
\cite{Mergul} concluded that the helicity of massless neutrinos is
not conserved when gravitational fields are present. On the
contrary, a calculation by Aldovrandi \emph{et alii} \cite{Ald}
based on linearized, scalar quantum gravity shows that helicity is
conserved for massless fermions.

Our purpose here is to re-examine some of these questions starting
from Hamiltonians that can be derived directly from the
covariant Dirac equation \cite{Singh1}.
Within the context of general relativity, comprehensive studies of
the Dirac equation were conducted by De Oliveira and Tiomno
\cite{DeOliveira} and Peres \cite{Peres}. Other authors introduce
inertial interactions by means of unitary transformations. This is
the approach taken by Bell and Leinaas \cite{Bell}. Their work
represents an extension of well known properties of the
Schr\"odinger equation to the relativistic regime, but cannot be
readily applied to gravitational fields. More recently, Hehl and
Ni \cite{Hehl} have derived a comprehensive Hamiltonian using
special relativity, while Obukhov \cite{Ob} has discussed some of
the limits of the most frequently used approximations.
Experimentally, the validity of the covariant Dirac equation in an
inertial-gravitational context finds support in the tests of the
Page-Werner \cite{Werner2} and Bonse-Wroblewski \cite{Bonse}
effects, and in the fact that spin-rotation coupling faithfully reproduces
the essential features of $g-2$ experiments without the
introduction of \emph{ad hoc} arguments \cite{PAP,Lamb}.
We use below the formalism of general relativity that treats both
inertial and gravitational fields in a unified way, but also avail
ourselves of solutions that are exact to first order in the weak
field approximation \cite{Cai1,Singh1}. We then develop suitable
low- and high-energy approximations.

The purpose of this paper is to study the time rate of change of
helicity and chirality for a massive, accelerated, charged spin-1/2
particle with total magnetic moment $\mu = (1+\kappa)\mu_{0}$,
where $\kappa \mu_{0}$ is the anomalous part of the magnetic
moment of the particle and $\mu_{0}$ is the Bohr magneton.
The paper is organized as follows. Section 2 describes the Dirac
Hamiltonian for a spin-1/2 particle under non-uniform acceleration
and rotation. In addition to the original Hamiltonian, low- and
high-energy approximations corresponding to non-relativistic and
ultra-relativistic particle motion are derived via the
Foldy-Wouthuysen (FW) \cite{Foldy} and Cini-Touschek (CT)
\cite{Cini} transformations, respectively. In Section 3, we
calculate the spin-flip transition rate for each of the
representations of the Hamiltonian. This is followed in Section 4
by the evaluation of the helicity operator's time evolution. Here
it is shown that a non-zero helicity precession emerges due solely
to the gravitational interactions found in the Berry's phase
approach \cite{Cai1,Cai2,Cai3,Berry}, even when the particle is
massless. Section 5 describes the chiral transition rate for a
spin-1/2 particle in accelerated motion, and is followed by the
conclusions in Section 6.

\section{Dirac Hamiltonian for an Accelerated Spin-1/2 Particle}
\setcounter{equation}{0}

\subsection{Original Representation}

Given the covariant Dirac equation\footnote{
Geometrized units of $c \ = \ 1$ are assumed throughout, where the
metric has signature $-2$.
Space-time indices are denoted by Greek characters and range from 0 to 3, while
spatial indices use Latin characters and range from 1 to 3.},
\be
\left[i \gamma^\mu (x) D_\mu - {m \over \hbar}\right]\psi (x) & = & 0,
\label{covDirac=}
\ee
where $m$ is the particle rest mass, $D_\mu \equiv \nabla_\mu
+ i \, \Gamma_\mu$ is the covariant derivative operator with
$\nabla_\mu$ the usual covariant derivative on index-labelled
tensors, and $\Gamma_\mu$ is the spinor connection,
we seek to derive a corresponding Dirac Hamiltonian in a general
co-ordinate frame.
The gamma matrices $\left\{ \gamma^\mu(x) \right\}$ satisfy
$\left\{\gamma^\mu (x), \gamma^\nu (x) \right\} = 2 \, g^{\mu
\nu}(x) $ and $D_\mu \, \gamma^\nu = 0$.
The metric is described as
\be
\vec{g} & = & \eta_{\hat{\mu}\hat{\nu}} \, \vec{e}^{\hat{\mu}} \otimes
\vec{e}^{\hat{\nu}},
\label{metric-ortho=}
\ee
where we use a set of orthonormal tetrads \cite{DeFelice}
$\left\{\vec{e}_{\hat{\mu}} \right\}$ and basis one-forms
$\left\{\vec{e}^{\hat{\mu}} \right\}$ labelled by indices with
carets and satisfying the condition $\left\langle
\vec{e}^{\hat{\mu}} , \vec{e}_{\hat{\nu}} \right\rangle =
\delta^{\hat{\mu}}{}_{\hat{\nu}}$ to define a local Lorentz frame.
With vierbein sets $\left\{ e^{\hat{\alpha}}{}_\mu \right\}$,
$\left\{ e^\mu{}_{\hat{\alpha}} \right\}$ satisfying
$\vec{e}^{\hat{\alpha}} = e^{\hat{\alpha}}{}_\beta \,
\vec{e}^\beta$ and $\vec{e}_{\hat{\alpha}} =
e^\beta{}_{\hat{\alpha}} \, \vec{e}_\beta$, such that
\be
e^{\hat{\alpha}}{}_\mu \, e^\mu{}_{\hat{\beta}} & = &
\delta^{\hat{\alpha}}{}_{\hat{\beta}}
\nn
e^{\mu}{}_{\hat{\alpha}} \, e^{\hat{\alpha}}{}_{\nu} & = &
\delta^\mu{}_\nu
\label{vierbein=}
\nl
g_{\mu \nu} & = & \eta_{\hat{\alpha}\hat{\beta}} \,
e^{\hat{\alpha}}{}_\mu \, e^{\hat{\beta}}{}_\nu,
\label{gtensor=}
\ee
we can relate the general metric to its Minkowski counterpart.
The spinor connection is then
\be
\Gamma_\mu & = & -{1 \over 4} \, \sigma^{\alpha \beta}(x) \,
\Gamma_{\alpha \beta \mu} \ = \
-{1 \over 4} \, \sigma^{\hat{\alpha} \hat{\beta}} \,
\Gamma_{\hat{\alpha} \hat{\beta} \hat{\mu}} \, e^{\hat{\mu}}{}_\mu,
\label{Gammadef=}
\ee
where $\sigma^{\hat{\alpha} \hat{\beta}} = {i \over 2}
[\gamma^{\hat{\alpha}}, \gamma^{\hat{\beta}}]$ are the
Minkowski space-time spin matrices and from the Cartan equation of
differential forms
\be
\d\vec{e}^{\hat{\mu}} + \Gamma^{\hat{\mu}}{}_{\hat{\beta} \hat{\alpha}} \,
\vec{e}^{\hat{\alpha}} \wedge \vec{e}^{\hat{\beta}} & = & 0,
\label{Cartan=}
\ee
we obtain $\Gamma_{\hat{\alpha} \hat{\beta} \hat{\mu}}$, the Ricci rotation
coefficients.
It is shown that, by arranging (\ref{covDirac=}) into a Schr\"{o}dinger form,
the Hamiltonian in general space-time co-ordinates is
$i \hbar \, \partial_0 \, \psi(x) = H \, \psi(x)$, where
\be
H & = & \left(g^{00}\right)^{-1} \, e^0{}_{\hat{\mu}} \left[
\gamma^{\hat{\mu}} \, m + e^j{}_{\hat{\nu}} \left(\eta^{\hat{\mu} \hat{\nu}}
- i \, \sigma^{\hat{\mu} \hat{\nu}} \right) \left(-i \hbar \, \nabla_j +
\hbar \, \Gamma_j \right) \right] + \hbar \, \Gamma_0.
\label{Hgeneral=}
\ee

The orthonormal tetrad \cite{Hehl} for a spin-1/2 particle under accelerated motion
with spatial rotational freedom is
\be
\vec{e}_{\hat{0}} & = & \left(1 + \vec{a} \cdot \vec{x} \right)^{-1}
\left[\partial_0 - \left(\vec{\omega} \times \vec{x} \right)^k \partial_k \right]
\nn
\vec{e}_{\hat{k}} & = & \partial_k
\label{ortho-tetrad=}
\ee
and the corresponding basis one-form is
\be
\vec{e}^{\hat{0}} & = & \left(1 + \vec{a} \cdot \vec{x} \right) \, \d x^0
\nn
\vec{e}^{\hat{k}} & = & \d x^k + \left(\vec{\omega} \times \vec{x} \right)^k \, \d x^0,
\label{1-form=}
\ee
where the three-acceleration $\vec{a}$ of the particle's frame and the rotation
$\vec{\omega}$ of its spatial triad are generated by external electromagnetic fields.
By introducing the electromagnetic potential and weak gravitational potential via the
covariant Berry's phase $\Phi_{\rm G}$ \cite{Singh1},
it follows from (\ref{Hgeneral=}) that the
Hamiltonian generated from (\ref{ortho-tetrad=}) and (\ref{1-form=}) is
\be
H & = & \left(1 + \vec{a} \cdot \vec{x} \right) \left[
\vec{\alpha} \cdot \vec{\pi} + m \, \beta + {\kappa e \hbar \over 2m} \beta
\left(i \vec{\alpha} \cdot \vec{E} - \vec{\sigma} \cdot \vec{B}\right) \right]
- {i \hbar \over 2} (\vec{\alpha} \cdot \vec{a})
\nn
&  &{}- \vec{\omega} \cdot (\vec{x} \times \vec{\pi}) - {\hbar \over 2} \,
\vec{\sigma} \cdot \vec{\omega} + e \, \varphi
+ \vec{\alpha} \cdot (\vec{\nabla} \Phi_{\rm G}) + (\nabla_0 \Phi_{\rm G}),
\label{H-accel=}
\ee
where $\vec{\pi} \equiv \vec{p} - e\vec{A}$ with momentum operator
$\vec{p}$ and electromagnetic vector potential $\vec{A}$,
the anomalous magnetic moment $ \kappa $ is inserted by hand, and the
covariant Berry's phase is
\be
\Phi_{\rm G} & = &
-\frac{1}{2}\int_{P}^{x}dz^{\lambda}\gamma_{\alpha\lambda}\left(z\right)p^{\alpha}+
\frac{1}{2}\int_{P}^{x}dz^{\lambda}
\left(\gamma_{\alpha\lambda,\beta}\left(z\right)-\gamma_{\beta\lambda,\alpha}\left(z\right)\right)
\left(x^{\alpha}-z^{\alpha}\right)p^{\beta},
\label{PH}
\ee
where $ p^{\mu}$ is the momentum eigenvalue of the free particle.

\subsection{Low- and High-Energy Approximations of the Hamiltonian}

It is possible to consider the precession of a spin-1/2 particle's
helicity state for non-relativistic and ultra-relativistic motion.
To do this, it is necessary that the Hamiltonian (\ref{H-accel=})
undergo a suitable transformation which appropriately describes
these energy limits. This is accomplished by using the
Foldy-Wouthuysen (FW) \cite{Foldy,Bjorken} and Cini-Touschek (CT)
\cite{Cini,Bose} transformations to respectively obtain the low-
and high-energy approximations of the Dirac Hamiltonian. For the
special case of the free-particle Hamiltonian $H_0 = m\beta +
\vec{\alpha} \cdot \vec{\pi}$, the resulting low- and high-energy
Hamiltonians $H_0^{\rm FW}$ and $H_0^{\rm CT}$ are well known.
However, these derivations assume use of a Cartesian co-ordinate
frame in performing the calculations. We want to generalize this
approach by assuming a general curvilinear co-ordinate frame such
that the results can then be applied to any orthogonal co-ordinate
system. To accomplish this, we begin with the unitary operator
$\exp\left(i S_{\rm FW/CT}\right)$, where
\be
S_{\rm FW/CT} & = & {i \over 2} \, \omega(q) \,
\beta \, {\left(\vec{\alpha} \cdot \vec{\pi}\right) \over |\vec{\pi}|},
\label{Sct=}
\ee
the momentum operator $\vec{p}$ in curvilinear co-ordinates is
\be
P^{\hat{\imath}} & = &  -i \hbar \, \nabla_{\hat{\imath}}
\ = \ -i \hbar \, {1 \over \lambda^{\hat{\imath}}(u)} \,
{\partial \over \partial u^{\hat{\imath}}},
\label{momentum-curve=}
\ee
with scale functions $\lambda^{\hat{\imath}}(u)$,
and $\omega(q)$ is a constraint function dependent on $q \equiv m/|\vec{\pi}|$
and to be determined.\footnote{We adopt the sign conventions used by Itzykson and Zuber
\cite{Itzykson} for the momentum operator and the three-dimensional Levi-Civita symbol,
where $\epsilon^{ijk} = \epsilon_{ijk}$ and $\epsilon^{123} \equiv +1$, noting that the
indices are raised and lowered with the Kronecker delta $\delta_{ij}$.}
Then
\be
H_0^{\rm FW/CT} & = & e^{i S_{\rm FW/CT}} \left[m \beta +
\vec{\alpha} \cdot \vec{\pi} \right] e^{-i S_{\rm FW/CT}}
\ = \
e^{2i S_{\rm FW/CT}} \left[m \beta + \vec{\alpha} \cdot \vec{\pi} \right].
\label{Hct1=}
\ee
By Taylor expansion, it is shown that
\be
e^{2i S_{\rm FW/CT}} & = &
1 - \omega \, \beta {\left(\vec{\alpha} \cdot \vec{\pi}\right) \over |\vec{\pi}|}
+ {1 \over 2!} \left[\omega \, \beta \,
{\left(\vec{\alpha} \cdot \vec{\pi}\right) \over |\vec{\pi}|} \right]^2
- {1 \over 3!} \left[\omega \, \beta \,
{\left(\vec{\alpha} \cdot \vec{\pi}\right) \over |\vec{\pi}|} \right]^3
+ \cdots
\label{exp1=}
\nl
\left[\omega \, \beta \,
{\left(\vec{\alpha} \cdot \vec{\pi}\right) \over |\vec{\pi}|} \right]^2
& = & - {\omega^2 \over |\vec{\pi}|^2} \left(\vec{\alpha} \cdot \vec{\pi}\right)^2
\ = \ - {\omega^2 \over |\vec{\pi}|^2} \left[\vec{\pi} \cdot \vec{\pi}
+ {i \over 2} \,  \epsilon_{ijk} \, \sigma^{\hat{\imath}} \, [P^{\hat{\jmath}}, P^{\hat{k}}]
- e \hbar \, \epsilon^i{}_{jk} \, \sigma^{\hat{k}} \,
\left(\nabla_{\hat{\imath}} A^{\hat{\jmath}} \right)\right],
\label{exp2=}
\ee
%
It is an important point to recognize that, for general curvilinear co-ordinates,
$\frac{i}{\hbar}[P^{\hat{\jmath}}, P^{\hat{k}}] \equiv N^{\hat{\jmath}{\hat{k}}} \neq 0$.
Therefore, it follows that we can identify a vector operator $\vec{R}$ with the form
\be
R^{\hat{k}} & = & {i \over 2\hbar} \, \epsilon_{ij}{}^k \, [P^{\hat{\imath}}, P^{\hat{\jmath}}]
\ = \ {1 \over 2} \, \epsilon_{ij}{}^k \, N^{\hat{\imath}{\hat{\jmath}}}
\nn
& = & \delta^k{}_1 \left[{1 \over \lambda^{\hat{3}}(u)}
\left({\partial \over \partial u^{\hat{3}}} \, \ln \lambda^{\hat{2}}(u) \right) P^{\hat{2}} -
{1 \over \lambda^{\hat{2}}(u)}
\left({\partial \over \partial u^{\hat{2}}} \, \ln \lambda^{\hat{3}}(u) \right) P^{\hat{3}}
\right]
\nn
&  &{} + \delta^k{}_2 \left[{1 \over \lambda^{\hat{1}}(u)}
\left({\partial \over \partial u^{\hat{1}}} \, \ln \lambda^{\hat{3}}(u) \right) P^{\hat{3}} -
{1 \over \lambda^{\hat{3}}(u)}
\left({\partial \over \partial u^{\hat{3}}} \, \ln \lambda^{\hat{1}}(u) \right) P^{\hat{1}}
\right]
\nn
&  &{} + \delta^k{}_3 \left[{1 \over \lambda^{\hat{2}}(u)}
\left({\partial \over \partial u^{\hat{2}}} \, \ln \lambda^{\hat{1}}(u) \right) P^{\hat{1}} -
{1 \over \lambda^{\hat{1}}(u)}
\left({\partial \over \partial u^{\hat{1}}} \, \ln \lambda^{\hat{2}}(u) \right) P^{\hat{2}}
\right].
\label{Rvec=}
\ee
Then, from (\ref{exp2=}), it is shown that
\be
\left[\omega \, \beta \, {\left(\vec{\alpha} \cdot \vec{\pi}\right) \over |\vec{\pi}|}\right]^2
& = & - \omega^2 \left[1 + {\hbar \over |\vec{\pi}|^2} \, \vec{\sigma} \cdot \vec{R} -
{e\hbar \over |\vec{\pi}|^2} \, \sigma^{\hat{k}} \, \epsilon^i{}_{jk} \,
\left(\nabla_{\hat{\imath}} A^{\hat{\jmath}} \right)\right] \ \equiv \ - \chi^2
\nn
\chi & \approx & \omega \left[1 + {\hbar \over 2|\vec{\pi}|^2} \, \vec{\sigma} \cdot \vec{R} -
{e\hbar \over 2|\vec{\pi}|^2} \, \sigma^{\hat{k}} \, \epsilon^i{}_{jk} \,
\left(\nabla_{\hat{\imath}} A^{\hat{\jmath}} \right)\right].
\label{(beta.alpha.p)^2=}
\ee
Given (\ref{(beta.alpha.p)^2=}), it is evident that $\vec{\sigma} \cdot \vec{R}$
resembles something like
a magnetic dipole term due to the curl of the electromagnetic vector potential.
However, this is interpreted as a purely co-ordinate-dependent effect due to the
choice of momentum states defined in a particular co-ordinate system.
For Cartesian co-ordinates, (\ref{Rvec=}) identically vanishes.
Therefore, by substituting into (\ref{exp1=}), we show that
\be
e^{2i S_{\rm FW/CT}} & \approx & \cos \chi - \sin \chi \left[\chi^{-1} \, \omega
\, \beta \, {\left(\vec{\alpha} \cdot \vec{\pi} \right) \over |\vec{\pi}|} \right]
\nn
& \approx & \cos \chi - \sin \chi \left[1 - {\hbar \over 2|\vec{\pi}|^2} \,
\vec{\sigma} \cdot \vec{R} + {e\hbar \over 2|\vec{\pi}|^2} \, \sigma^{\hat{k}} \, \epsilon^i{}_{jk} \,
\left(\nabla_{\hat{\imath}} A^{\hat{\jmath}} \right)\right]
\beta \, {\left(\vec{\alpha} \cdot \vec{\pi} \right) \over |\vec{\pi}|},
\label{exp3=}
\ee
and
\be
H_0^{\rm FW/CT} & \approx & \left[\cos \chi + q \, \sin \chi
\left[1 - {\hbar \over 2|\vec{\pi}|^2} \,
\vec{\sigma} \cdot \vec{R} + {e\hbar \over 2|\vec{\pi}|^2} \, \sigma^{\hat{k}} \, \epsilon^i{}_{jk} \,
\left(\nabla_{\hat{\imath}} A^{\hat{\jmath}} \right)\right]
\right]\left(\vec{\alpha} \cdot \vec{\pi} \right)
\nn
&  &{} + |\vec{\pi}| \left[q \, \cos \chi - \sin \chi \left[1 + {\hbar \over 2|\vec{\pi}|^2} \,
\vec{\sigma} \cdot \vec{R} - {e\hbar \over 2|\vec{\pi}|^2} \, \sigma^{\hat{k}} \, \epsilon^i{}_{jk} \,
\left(\nabla_{\hat{\imath}} A^{\hat{\jmath}} \right)\right]
\right] \beta.
\label{H_free}
\ee

By setting the first coefficient of (\ref{H_free}) to zero, we obtain the low-energy
Hamiltonian, $H_0^{\rm FW}$, which amounts to solving for $\omega(q)$.
It is straightforward to show that
\be
\omega(q) & \approx & \left[1 - {\hbar \over 2|\vec{\pi}|^2} \,
\vec{\sigma} \cdot \vec{R} + {e\hbar \over 2|\vec{\pi}|^2} \, \sigma^{\hat{k}} \, \epsilon^i{}_{jk} \,
\left(\nabla_{\hat{\imath}} A^{\hat{\jmath}} \right)\right]
\tan^{-1} \left[-{1 \over q}
\left[1 + {\hbar \over 2|\vec{\pi}|^2} \,
\vec{\sigma} \cdot \vec{R}
- {e\hbar \over 2|\vec{\pi}|^2} \, \sigma^{\hat{k}} \, \epsilon^i{}_{jk} \,
\left(\nabla_{\hat{\imath}} A^{\hat{\jmath}} \right)\right] \right]
\nn
& \approx & -{1 \over q} \ = \ -{|\vec{\pi}| \over m} \ \ll \ 1,
\label{omegaFW=}
\ee
with the result that
\be
H_0^{\rm FW} & = &  \left[m + {1 \over 2m} \, \vec{\pi} \cdot \vec{\pi}
+ {\hbar \over 2m} \, \vec{\sigma} \cdot \vec{R}
- {e \hbar \over 2m} \, \sigma^{\hat{k}} \, \epsilon^i{}_{jk} \left(\nabla_{\hat{\imath}}
A^{\hat{\jmath}} \right) \right] \beta.
\label{H_freeFW}
\ee
In a local Cartesian frame, (\ref{H_freeFW}) becomes
\be
H_0^{\rm FW} & = & \left[m + {1 \over 2m} \, \vec{\pi} \cdot \vec{\pi}
- {e \hbar \over 2m} \, \vec{\sigma} \cdot \vec{B} \right] \beta,
\label{H_freeFW_cart}
\ee
where the last term is the familiar magnetic dipole moment.\footnote{We must note that
$\epsilon^i{}_{jk} \left(\nabla_{\hat{\imath}} A^{\hat{\jmath}} \right)$ is {\em not} the
kth component of the curl of $\vec{A}$ in curvilinear co-ordinates \cite{Riley}.
As well, all summations involving gradients throughout this paper, as defined by
(\ref{momentum-curve=}),
are interpreted such that, for example, $\nabla_{\hat{k}} \nabla^{\hat{k}} \varphi =
-\left[\nabla_{\hat{1}}\nabla_{\hat{1}} + \nabla_{\hat{2}}\nabla_{\hat{2}} +
\nabla_{\hat{3}}\nabla_{\hat{3}}\right] \varphi$.}

Similarly, setting the second coefficient of (\ref{H_free}) to zero leads to the high-energy
approximation, $H_0^{\rm CT}$.
Following the same procedure, we show that
\be
\omega(q) & \approx & \left[1 - {\hbar \over 2|\vec{\pi}|^2} \,
\vec{\sigma} \cdot \vec{R} + {e\hbar \over 2|\vec{\pi}|^2} \, \sigma^{\hat{k}} \, \epsilon^i{}_{jk} \,
\left(\nabla_{\hat{\imath}} A^{\hat{\jmath}} \right)\right]
\tan^{-1} \left[q
\left[1 - {\hbar \over 2|\vec{\pi}|^2} \, \vec{\sigma} \cdot \vec{R}
+ {e\hbar \over 2|\vec{\pi}|^2} \, \sigma^{\hat{k}} \, \epsilon^i{}_{jk} \,
\left(\nabla_{\hat{\imath}} A^{\hat{\jmath}} \right)\right] \right]
\nn
& \approx & q
\left[1 - {\hbar \over |\vec{\pi}|^2} \, \vec{\sigma} \cdot \vec{R}
+ {e\hbar \over |\vec{\pi}|^2} \, \sigma^{\hat{k}} \, \epsilon^i{}_{jk} \,
\left(\nabla_{\hat{\imath}} A^{\hat{\jmath}} \right)\right] \ \ll \ 1,
\label{omegaCT=}
\ee
and
\be
H_0^{\rm CT} & \approx & \left[\sqrt{|\vec{\pi}|^2 + m^2} - {q^3 \over \sqrt{1 + q^2}}
\, {\hbar \over 2m} \left[\vec{\sigma} \cdot \vec{R}
- e \, \sigma^{\hat{k}} \, \epsilon^i{}_{jk} \left(\nabla_{\hat{\imath}}
A^{\hat{\jmath}} \right)\right] \right]
{\left(\vec{\alpha} \cdot \vec{\pi}\right)\over |\vec{\pi}|}.
\label{H_freeCT}
\ee
Both (\ref{H_freeFW}) and (\ref{H_freeCT}) show that a small energy shift due to $\vec{R}$
emerges from applying the FW and CT transformations to $H_0$.

Having now obtained the means to extend the FW and CT
transformations for a general curvilinear co-ordinate frame, we
proceed to derive the low- and high-energy approximations of the
Dirac Hamiltonian (\ref{H-accel=}) for a non-uniformly accelerated
spin-1/2 particle. Retaining all Hermitian terms up to order
$1/m^2$, it is therefore shown that\footnote{For notational
purposes, the indices are left uncaretted for the remainder of this paper.}
\be
H_{\rm FW} & \approx &
\left(1 + \vec{a} \cdot \vec{x} \right) \left[m + {1 \over 2m} \, \vec{\pi} \cdot \vec{\pi}
+ {\hbar \over 2m} \, \vec{\sigma} \cdot \vec{R}
- {e \hbar \over 2m} \, \sigma^k \, \epsilon^i{}_{jk} \left(\nabla_i A^j \right) \right] \beta
- {\kappa e \hbar \over 2m} \left( \vec{\sigma} \cdot \vec{B} \right)\beta
\nn
&  & {} + {\hbar \over 4m} \left[\vec{\sigma} \cdot \left(\vec{\nabla}\left(\vec{a} \cdot \vec{x}\right)
\times \vec{\pi} \right) - \hbar \left(\nabla_k a^k \right) + {\hbar \over 2} \,
\vec{\nabla} \cdot \vec{\nabla}\left(\vec{a} \cdot \vec{x}\right) \right]\beta
+ \left[{1 \over m} \, (\vec{\nabla} \Phi_{\rm G}) \cdot \vec{\pi} +
{\hbar \over 2m} \,  \sigma^k \, \epsilon^{ij}{}_k
\left(\nabla_i \, \nabla_j \Phi_{\rm G} \right)\right] \beta
\nn
&  &{} +
{e \hbar^2 \over 8m^2} \left[\vec{\nabla} \cdot \vec{\nabla} \varphi - 2 \kappa \left(\nabla_k E^k \right) \right]
+ {e \hbar \over 4m^2} \, \vec{\sigma} \cdot \left[\left(\vec{\nabla} \varphi \times \vec{\pi} \right)
- 2 \kappa \left(\vec{E} \times \vec{\pi}\right) \right]
\nn
&  &{}
- {\hbar^2 \over 8m^2} \left[ \vec{\nabla} \cdot \vec{\nabla} \left[\left(\vec{\omega} \times \vec{x} \right)
\cdot \vec{\pi} \right] + \nabla_k \left(\vec{\omega} \times \vec{\pi} \right)^k \right] -
{\hbar \over 4m^2} \,
\vec{\sigma} \cdot \left[\left(\vec{\omega} \times \vec{\pi} \right) \times \vec{\pi} \right]
\nn
&  &{} + {\hbar^2 \over 8m^2} \, \vec{\nabla} \cdot \vec{\nabla}(\nabla_0 \Phi_{\rm G})
+ {\hbar \over 4m^2} \,
\vec{\sigma} \cdot \left[\vec{\nabla}(\nabla_0 \Phi_{\rm G}) \times \vec{\pi} \right]
- \vec{\omega} \cdot (\vec{x} \times \vec{\pi}) - {\hbar \over 2} \, \vec{\sigma} \cdot \vec{\omega}
+ e \, \varphi +  (\nabla_0 \Phi_{\rm G}).
\label{H-FW=}
\ee
It is important to emphasize that terms such as $\vec{\nabla} \cdot \vec{\nabla} \varphi$ and
$\sigma^k \, \epsilon^{ij}{}_k \left( \nabla_i \, \nabla_j \Phi_{\rm G} \right)$ become
$\vec{\nabla}^2 \varphi$ and $\vec{\sigma} \cdot \left[\vec{\nabla} \times \vec{\nabla} \Phi_{\rm G}\right]
= 0$ {\em only} for
Cartesian co-ordinates, due to scale functions of $\lambda^k(u) = 1$ in the definition of the
gradient operator.
Clearly, this also implies that $\vec{R} = 0$ under the same circumstance.
Again, retaining only the leading-order Hermitian terms, we show that the high-energy
approximation of the Hamiltonian is
\be
H_{\rm CT} & \approx &
\left(1 + \vec{a} \cdot \vec{x} \right) \left[
\left[\sqrt{|\vec{\pi}|^2 + m^2} - {q^3 \over \sqrt{1 + q^2}}
\, {\hbar \over 2m} \, \vec{\sigma} \cdot \vec{R'} \right]
{\left(\vec{\alpha} \cdot \vec{\pi}\right)\over |\vec{\pi}|}
+ {\kappa e \hbar \over 2m} \beta
\left(i \vec{\alpha} \cdot \vec{E} - \vec{\sigma} \cdot \vec{B}\right) \right]
\nn
&  &{} - \vec{\omega} \cdot (\vec{x} \times \vec{\pi}) -
{\hbar \over 2} \, \vec{\sigma} \cdot \vec{\omega} + e \, \varphi + \vec{\alpha} \cdot
\vec{\nabla} \Phi_{\rm G} + (\nabla_0 \Phi_{\rm G})
\nn
&  &{} + {q \over 2|\vec{\pi}|}
\left(1 + \vec{a} \cdot \vec{x} \right) {\kappa e \hbar \over 2m} \left[\hbar \left(\nabla_k E^k \right) +
2 \, \vec{\sigma} \cdot \left(\vec{E} \times \vec{\pi} \right) - {2\hbar \over |\vec{\pi}|^2} \,
\vec{R'} \cdot \left(\vec{E} \times \vec{\pi} \right) -
2 \left(1 - {\hbar \over |\vec{\pi}|^2} \, \vec{\sigma} \cdot \vec{R'} \right) \vec{B} \cdot \vec{\pi} \right]
\nn
&  &{} + {q \over 2|\vec{\pi}|} \left[-{\hbar \over |\vec{\pi}|} \sqrt{|\vec{\pi}|^2 + m^2}
\left(1 - {\hbar \over |\vec{\pi}|^2} \, \vec{\sigma} \cdot \vec{R'} \right) \sigma^k \, \epsilon^i{}_{jk} \,
\nabla_i \left(\vec{a} \cdot \vec{x} \right) \pi^j + {\hbar^2 \over |\vec{\pi}|^2} \,
\vec{\alpha} \cdot \left[\vec{R'} \times \vec{\nabla} \left(\left(\vec{\omega} \times \vec{x} \right) \cdot
\vec{\pi} \right) \right]
\right.
\nn
&  &{} - {\hbar^2 \over 2} \, \alpha^k \, \epsilon^i{}_{jk} \left(\nabla_i \omega^j \right)
+ {\hbar^2 \over |\vec{\pi}|^2} \left[\left(\vec{R'} \cdot \vec{\omega} \right) \vec{\alpha} \cdot \vec{\pi}
- \alpha^j \, R'^k \, \omega_j \, \pi_k \right]
\nn
&  &{} + \left. 2 \left(1 - {\hbar \over |\vec{\pi}|^2} \, \vec{\sigma} \cdot \vec{R'} \right)
\left[ \vec{\nabla} \Phi_{\rm G} \cdot \vec{\pi} - {\hbar^2 \over 2} \left( \nabla_k a^k \right)
\right] \right]\beta,
\label{H-CT=}
\ee
where $R'^k = R^k - e \, \epsilon^{ki}{}_j \left(\nabla_i A^j\right)$.

\section{Spin-Flip Transition Rate}
\setcounter{equation}{0}

The helicity operator $h$ is given by
\be
h & \equiv & {\vec{\sigma} \cdot \vec{\pi} \over |\vec{\pi}|},
\label{helicity=}
\ee
and the spin-flip transition rate is
\be
\qquad {d h \over d t} \ = \
{i \over \hbar} \, \left[H, h\right].
\label{hdot-general=}
\ee
From (\ref{H-accel=}) and (\ref{helicity=}), it is straightforward to show
that the helicity precession operator is
\be
|\vec{\pi}| \, {d h \over d t} & = &
\vec{\sigma} \cdot \left[ \left(\vec{\omega} + i \, \vec{a}
\right) \times \vec{\pi} - {\hbar \over 2} \, \epsilon^k{}_{lm}
\left[\left(\nabla_k a^l \right) - i \left( \nabla_k \omega^l \right)\right]
\hat{\vec{x}}^m + \left(\vec{\omega} \times \vec{x} \right) \times
\vec{R}
+ \epsilon_{klm} \left[\vec{\nabla} \left(\omega^l \, x^m\right) \right]
\pi^k \right.
\nn
&  &{} - \left. e \left(\vec{\omega}
\times \vec{x} \right)^k \left[\left(\nabla_k A_l \right) - \left(\nabla_l A_k \right)\right]
\hat{\vec{x}}^l  - e \, \vec{\nabla} \varphi - \vec{\nabla}(\nabla_0 \Phi_{\rm G}) \right]
\nn
&  &{} -
\vec{\sigma} \cdot \left[{2 \over \hbar} \, \vec{\nabla} \Phi_{\rm G} \times \vec{\pi}
+ i \, \vec{\nabla} \left(\vec{a} \cdot \vec{x}\right) \times \vec{\pi}
+ i \, \epsilon^{kl}{}_m \left( \nabla_k \nabla_l \Phi_{\rm G} \right) \hat{\vec{x}}^m
\right] \gamma^5
\nn
&  &{} + \vec{\sigma} \cdot \left[- m \, \vec{\nabla} \left(\vec{a}
\cdot \vec{x}\right) + {\kappa e \hbar \over 2m} \left[{2 \over
\hbar} \left(1 + \vec{a} \cdot \vec{x} \right) \vec{B} \times
\vec{\pi}
+ i \left(1 + \vec{a} \cdot \vec{x} \right) \epsilon^k{}_{lm} \left( \nabla_k  B^l \right)
\hat{\vec{x}}^m
+ i \, \vec{\nabla} \left(\vec{a} \cdot \vec{x}\right) \times \vec{B}
\right] \right] \beta
\nn
&  &{} - {\kappa e \hbar \over 2m} \,
\vec{\sigma} \cdot \left[\vec{\nabla} \left(\vec{a} \cdot
\vec{x}\right) \times \vec{E} + \left(1 + \vec{a} \cdot \vec{x}
\right) \, \epsilon^k{}_{lm} \left( \nabla_k E^l \right) \hat{\vec{x}}^m
- {2i \over \hbar} \left(1 + \vec{a} \cdot \vec{x} \right) \vec{E}
\times \vec{\pi} \right] \gamma^5 \, \beta
\nn
&  &{} + {\hbar \over 2} \left[\left(\nabla_k \omega^k\right) + i \left( \nabla_k a^k
\right) \right] -
\left[\vec{\nabla} \left(\vec{a} \cdot \vec{x}\right) \cdot
\vec{\pi} + \vec{\nabla} \cdot \vec{\nabla} \Phi_{\rm G} \right]\gamma^5
\nn
&  &{} + {\kappa e \hbar \over 2m}
\left[\vec{\nabla} \left(\vec{a} \cdot \vec{x}\right) \cdot
\vec{B} + \left(1 + \vec{a} \cdot \vec{x} \right) \left(\nabla_k B^k \right)
\right] \beta + {i \kappa e \hbar \over 2m} \left[\vec{\nabla}
\left(\vec{a} \cdot \vec{x}\right) \cdot \vec{E} + \left(1 +
\vec{a} \cdot \vec{x} \right) \left(\nabla_k E^k \right)\right] \gamma^5 \,
\beta,
\label{hdot=}
\ee
where $\hat{\vec{x}}^m$ is a unit vector of the spatial triad corresponding to the general
curvilinear co-ordinate system.

It is a straightforward process to evaluate $d h_{\rm FW} / d t$ and
$d h_{\rm CT} / d t$ using the Hamiltonians (\ref{H-FW=}) and (\ref{H-CT=}).
However, to do this first requires that $h$ gets converted into an equivalent form
using the FW and CT transformations, respectively, given that
\be
{d \over d t} \, h_{\rm FW/CT} & = &
e^{i S_{\rm FW/CT}} \, {d h \over d t} \, e^{-i S_{\rm FW/CT}}
\ = \ e^{i S_{\rm FW/CT}} {i \over \hbar} \left[H, h \right] e^{-i S_{\rm FW/CT}} \ = \
{i \over \hbar} \left[H_{\rm FW/CT},  h_{\rm FW/CT} \right].
\label{Heisenberg=}
\ee
Therefore, the transformed versions of the helicity operator are
\be
h_{\rm FW} & \approx & {1 \over  |\vec{\pi}|} \left\{ \vec{\sigma} \cdot \vec{\pi} -
{1 \over 8m^2} \left[{\cal O}, \left[{\cal O}, \vec{\sigma} \cdot \vec{\pi} \right] \right] \right\}
\nn
h_{\rm CT} & \approx & {\vec{\sigma} \cdot \vec{\pi} \over |\vec{\pi}|} \ = \ h,
\label{h-convert=}
\ee
where ${\cal O}$ is the set of odd operators which comprise the original Hamiltonian.

For the low-energy approximation, it is shown that the spin-flip transition rate is
\be
{d \over d t} \, h_{\rm FW} & \approx & {i \over \hbar} \left[H_{\rm FW},  h \right] +
{i \over \hbar} \left[H_{\rm FW},  h_1 \right],
\ee
where $h_1 = - \left(1/8m^2\right) \left[{\cal O}, \left[{\cal O}, h \right] \right]$, and
\be
\lefteqn{|\vec{\pi}| \, {i \over \hbar} \left[H_{\rm FW},  h \right] \ \approx \
-\left(1 + \vec{a} \cdot \vec{x} \right)
\left[{1 \over 2m} \, \vec{\sigma} \cdot \vec{\nabla} \left(\vec{\pi} \cdot \vec{\pi}\right)
+ {\hbar \over 2m} \left[\left(\nabla_k R'^k\right) + i \, \sigma^k \, \epsilon^i{}_{jk} \left(\nabla_i R'^j \right) \right]
+ {1 \over m} \, \vec{\sigma} \cdot \left(\vec{R'} \times \vec{\pi} \right) \right]\beta
 }
\nn
&&{} - \vec{\sigma} \cdot \vec{\nabla}\left(\vec{a} \cdot \vec{x}\right)
\left[m + {1 \over 2m} \, \vec{\pi} \cdot \vec{\pi} \right]\beta +
{\kappa e \over m} \, \vec{\sigma} \cdot \left(\vec{B} \times \vec{\pi}\right) \beta +
{\kappa e \hbar \over 2m} \left[\left(\nabla_k B^k\right) + i \, \sigma^k \,
\epsilon^i{}_{jk} \left( \nabla_i B^j \right)\right]\beta
\nn
&&{}  - {\hbar \over 2m}
\left[\vec{\nabla}\left(\vec{a} \cdot \vec{x}\right) \cdot \vec{R'} +
\sigma^k \, \epsilon^{ij}{}_k \left( \nabla_i \, \nabla_j \Phi_{\rm G}\right)
+ i \, \vec{\sigma} \cdot
\left[\vec{\nabla}\left(\vec{a} \cdot \vec{x}\right) \times \vec{R'} \right] +
i \, \sigma^j \, \nabla^i \left[\nabla_i \nabla_j - \nabla_j \nabla_i \right]\Phi_{\rm G}
\right] \beta
\nn
&&{}  - {1 \over 2m}
\left[\vec{\sigma} \cdot \left[ \left(\vec{\nabla}\left(\vec{a} \cdot \vec{x}\right)
\times \vec{\pi} \right)\times \vec{\pi} + 2 \, \vec{\nabla}(\vec{\nabla} \Phi_{\rm G} \cdot \vec{\pi}) \right] +
2 \, \sigma^j \left(
\left[\nabla_i \nabla_j - \nabla_j \nabla_i \right]\Phi_{\rm G} \right) \pi^i \right] \beta
\nn
&&{} + {\hbar \over 4m} \left[-\left[\nabla_k \left(\vec{\nabla}
\left(\vec{a} \cdot \vec{x}\right)\times \vec{\pi} \right)^k \right]
+ \hbar \, \vec{\sigma} \cdot \vec{\nabla}\left(\nabla_k a^k \right) -
{\hbar \over 2} \, \vec{\sigma} \cdot \vec{\nabla}
\left(\vec{\nabla} \cdot \vec{\nabla}\left(\vec{a} \cdot \vec{x}\right) \right) -
i \, \sigma^k \, \epsilon^i{}_{jk} \left[ \nabla_i \left(\vec{\nabla}\left(\vec{a} \cdot \vec{x}\right)
\times \vec{\pi} \right)^j \right] \right]\beta
\nn
&&{} + {\hbar^2 \over 8m^2} \, \vec{\sigma} \cdot \vec{\nabla}
\left[\vec{\nabla} \cdot \vec{\nabla} \left[\left(\vec{\omega} \times \vec{x} \right)
\cdot \vec{\pi} \right] + \left[\nabla_k \left(\vec{\omega} \times \vec{\pi} \right)^k \right]
+ e \left[ \nabla_k  \left(2 \kappa \, E^k + \nabla^k \varphi \right)\right] -
\vec{\nabla} \cdot \vec{\nabla}(\nabla_0 \Phi_{\rm G}) \right]
\nn
&&{} +
{\hbar \over 4m^2} \left[\left[\nabla_k
\left[\left(\vec{\omega} \times \vec{\pi} \right) \times \vec{\pi}
+ e \left(2\kappa \, \vec{E} - \vec{\nabla} \varphi\right)\times \vec{\pi} -
\vec{\nabla}(\nabla_0 \Phi_{\rm G}) \times \vec{\pi} \right]^k \right] \right.
\nn
&&{} + \left. i \, \sigma^k \, \epsilon^i{}_{jk} \left[ \nabla_i
\left[\left(\vec{\omega} \times \vec{\pi} \right) \times \vec{\pi}
+ e \left(2\kappa \, \vec{E} - \vec{\nabla} \varphi\right)\times \vec{\pi} -
\vec{\nabla}(\nabla_0 \Phi_{\rm G}) \times \vec{\pi} \right]^j \right] \right]
\nn
&&{} +
{1 \over 2m^2} \, \sigma^k \, \epsilon_{ijk} \left[\left(\vec{\omega} \times \vec{\pi} \right) \times \vec{\pi}
+ e \left(2\kappa \, \vec{E} - \vec{\nabla} \varphi\right)\times \vec{\pi} -
\vec{\nabla}(\nabla_0 \Phi_{\rm G}) \times \vec{\pi}\right]^i \, \pi^j
\nn
&&{} +
\vec{\sigma} \cdot \left[\left(\vec{\omega} \times \vec{x} \right) \times \vec{R} \right] -
e \, \sigma^j \, \left(\vec{\omega} \times \vec{x} \right)^i
\left[\left(\nabla_i A_j\right) - \left(\nabla_j A_i\right) \right]
+ {\hbar \over 2} \left[\left(\nabla_k \omega^k\right) + i \, \sigma^k \, \epsilon^i{}_{jk}
\left(\nabla_i \omega^j \right) \right]
\nn
&&{} + \vec{\sigma} \cdot \left(\vec{\omega} \times \vec{\pi} \right) +
\epsilon_{klm}
\left[\vec{\sigma} \cdot \vec{\nabla} \left(\omega^l \, x^m\right)\right]\pi^k -
\vec{\sigma} \cdot \vec{\nabla} \left(e \varphi +  (\nabla_0 \Phi_{\rm G})\right)
\label{hdot-FW0=}
\ee
\be
|\vec{\pi}| \, {i \over \hbar} \left[H_{\rm FW},  h_1 \right] & \approx &
- {\hbar^2 \over 8m^2} \left[
\sigma^k \left(\vec{\omega} \times \vec{x} \right)^j \left[\nabla_j \nabla_k
\left(\nabla_i \pi^i \right)\right] -
\vec{\sigma} \cdot \left[\vec{\omega} \times \vec{\nabla} \left(\nabla_k \pi^k \right) \right] \right]
\nn
& &{} + {\hbar^2 \over 4m^2} \left[
-\left(\vec{\omega} \times \vec{x} \right) \cdot \vec{\nabla}\left[
\sigma^k \, \nabla^j \left[N_{jk} + e \left(\nabla_j A_k\right) -
e \left(\nabla_k A_j\right) \right] \right] \right.
\nn
& &{} - \left. \sigma^k \, \epsilon_{ijk} \, \omega^i \left[\left(\nabla_m N^{mj}\right) +
e \left(\nabla_m \nabla^m A^j\right) - e \left(\nabla_m \nabla^j A^m \right) \right] \right]
\nn
& &{} + {i \hbar^2 \over 4m^2} \left[\left(\vec{\omega} \times \vec{x} \right) \cdot \vec{\nabla}
\left[\nabla_k \left[R^k - e \, \epsilon^{ik}{}_j \left(\nabla_i A^j \right)\right] \right] +
\left[N^{mn} + e \left(\nabla^m A^n\right) - e \left(\nabla^n A^m \right)\right]
\left(\nabla_m \omega_n \right)  \right.
\nn
& &{} - \left. i \, \sigma^k \, \epsilon_{ijk} \left[N^{mi} +
e \left(\nabla^m A^i\right) - e \left(\nabla^i A^m \right)\right] \left(\nabla_m \omega^j \right) \right]
\nn
& &{} - {i \hbar^2 \over 2m^2} \left[\sigma^k \left(\vec{\omega} \times \vec{x} \right) \cdot \vec{\nabla}
\left[
\left[N_{mk} + e \left(\nabla_m A_k\right) - e \left(\nabla_k A_m \right)
\right]\pi^m \right] \right.
\nn
& &{} - \left. \sigma^k \, \epsilon_{ijk} \, \omega^i \left[N^{mj} +
e \left(\nabla^m A^j\right) - e \left(\nabla^j A^m \right) \right]\pi_m \right].
\label{hdot-FW1=}
\ee
In a similar fashion, for $H_{\rm CT} = H_0 + q \, H_1$, the spin-flip transition rate for an
ultra-relativistically moving spin-1/2 particle follows from (\ref{Heisenberg=}), where
\be
\lefteqn{|\vec{\pi}| \, {i \over \hbar} \left[H_0,  h \right] \ = \
- {1 \over |\vec{\pi}|} \, \sqrt{|\vec{\pi}|^2 + m^2}
\left[\vec{\nabla}\left(\vec{a} \cdot \vec{x}\right) \cdot \vec{\pi} + i \, \vec{\sigma} \cdot
\left[\vec{\nabla}\left(\vec{a} \cdot \vec{x}\right) \times \vec{\pi} \right] \right]\gamma^5}
\nn
&&{} + {q^3 \over \sqrt{1 + q^2}} \, {\hbar \over 2m} \, {1 \over |\vec{\pi}|}
\left[\left[\vec{\sigma} \cdot \vec{\nabla}\left(\vec{a} \cdot \vec{x}\right)\right] \vec{R'} \cdot \vec{\pi} +
i \left[ \vec{\nabla}\left(\vec{a} \cdot \vec{x}\right)\right] \cdot \left(\vec{R'} \times \vec{\pi}\right) -
\vec{\sigma} \cdot \left[\vec{\nabla}\left(\vec{a} \cdot \vec{x}\right) \times
\left(\vec{R'} \times \vec{\pi}\right) \right] \right] \gamma^5
\nn
&&{}
+ \left(1 + \vec{a} \cdot \vec{x} \right)
\left[
{q^3 \over \sqrt{1 + q^2}} \, {\hbar \over 2m} \, {1 \over |\vec{\pi}|}
\left[\left[\vec{\sigma} \cdot \vec{\nabla}\left(\vec{R'} \cdot \vec{\pi}\right)\right] +
{2i \over \hbar} \, \vec{\sigma} \cdot \left[\left(\vec{R'} \times \vec{\pi}\right) \times \vec{\pi}\right]
\right. \right.
\nn
&&{} + \left. \left.
i \left[ \nabla_k \left(\vec{R'} \times \vec{\pi}\right)^k\right] - \sigma^k \,
\epsilon^i{}_{jk} \left[ \nabla_i
\left(\vec{R'} \times \vec{\pi}\right)^j \right]\right] \right]\gamma^5
\nn
&&{} + {\kappa e \hbar \over 2m}
\vec{\sigma} \cdot \left[{2 \over \hbar} \left(1 + \vec{a} \cdot \vec{x} \right)
\vec{B} \times \vec{\pi} + i \left(1 + \vec{a} \cdot \vec{x} \right) \epsilon^k{}_{lm} \left(
\nabla_k  B^l \right) \hat{\vec{x}}^m
+ i \, \vec{\nabla} \left(\vec{a} \cdot \vec{x}\right) \times \vec{B} \right] \beta
\nn
&&{} - {\kappa e \hbar \over 2m} \, \vec{\sigma} \cdot
\left[\vec{\nabla} \left(\vec{a} \cdot \vec{x}\right) \times \vec{E} +
\left(1 + \vec{a} \cdot \vec{x} \right) \, \epsilon^k{}_{lm} \left( \nabla_k E^l \right) \hat{\vec{x}}^m -
{2i \over \hbar} \left(1 + \vec{a} \cdot \vec{x} \right) \vec{E} \times \vec{\pi}
\right] \gamma^5 \, \beta
\nn
&&{} + {\kappa e \hbar \over 2m} \left[\vec{\nabla} \left(\vec{a} \cdot \vec{x}\right) \cdot \vec{B}
+ \left(1 + \vec{a} \cdot \vec{x} \right)\left(\nabla_k B^k \right)\right] \beta +
{i \kappa e \hbar \over 2m} \left[\vec{\nabla} \left(\vec{a} \cdot \vec{x}\right) \cdot \vec{E}
+ \left(1 + \vec{a} \cdot \vec{x} \right) \left(\nabla_k E^k \right)\right] \gamma^5 \, \beta
\nn
&&{} - \vec{\sigma} \cdot \left[{2 \over \hbar} \, \vec{\nabla} \Phi_{\rm G} \times \vec{\pi} +
i \, \epsilon^{kl}{}_m \left( \nabla_k \nabla_l  \Phi_{\rm G} \right) \hat{\vec{x}}^m \right]\gamma^5
+ {\hbar \over 2} \left( \nabla_k \omega^k \right) - \vec{\nabla} \cdot \vec{\nabla}\Phi_{\rm G} \, \gamma^5
\nn
&&{} + \vec{\sigma} \cdot \left[\vec{\omega} \times \vec{\pi} + {i\hbar \over 2} \, \epsilon^k{}_{lm}
\left(\nabla_k \omega^l \right) \hat{\vec{x}}^m + \left(\vec{\omega} \times \vec{x} \right) \times \vec{R}
+ \epsilon_{klm} \left[\vec{\nabla}\left(\omega^l \, x^m \right)\right] \pi^k \right.
\nn
&&{} - \left.
e \left(\vec{\omega} \times \vec{x} \right)^k \left[\left(\nabla_k A_l\right) -
\left(\nabla_l A_k\right) \right]\hat{\vec{x}}^l
- e \vec{\nabla} \varphi - \vec{\nabla}(\nabla_0 \Phi_{\rm G}) \right]
\label{hdot-CT0=}
\nl
\lefteqn{|\vec{\pi}| \, {i \over \hbar} \left[H_1,  h \right] \ = \
- {1 \over |\vec{\pi}|} \, {\kappa e \hbar \over 2m} \left[
\vec{\nabla} \left(\vec{a} \cdot \vec{x} \right) \cdot \left(\vec{E} \times \vec{\pi} \right)
+ i \, \vec{\sigma} \cdot \left[\vec{\nabla} \left(\vec{a} \cdot \vec{x} \right) \times
\left(\vec{E} \times \vec{\pi} \right) \right] - \left[\vec{\sigma} \cdot
\vec{\nabla} \left(\vec{a} \cdot \vec{x} \right)\right] \vec{B} \cdot \vec{\pi} \right]
}
\nn
&&{} +
{1 \over |\vec{\pi}|} \left(1 + \vec{a} \cdot \vec{x} \right)
 {\kappa e \hbar \over 2m} \left[
- \left[\left[\nabla_k \left(\vec{E} \times \vec{\pi} \right)^k \right]
+ i \, \sigma^k \, \epsilon^i{}_{jk} \left[
\nabla_i \left(\vec{E} \times \vec{\pi} \right)^j \right]\right] - {2 \over \hbar} \,
\vec{\sigma} \cdot \left[\left(\vec{E} \times \vec{\pi} \right) \times \vec{\pi} \right] +
\vec{\sigma} \cdot \vec{\nabla} \left(\vec{B} \cdot \vec{\pi} \right) \right]
\nn
&&{}
 + {1 \over 2|\vec{\pi}|} \left[{\hbar \over |\vec{\pi}|} \sqrt{|\vec{\pi}|^2 + m^2}
\left[\left[\nabla_k \left(\vec{\nabla}\left(\vec{a} \cdot \vec{x}\right) \times \vec{\pi}\right)^k \right]
+ {2 \over \hbar} \, \vec{\sigma} \cdot \left[
\left(\vec{\nabla}\left(\vec{a} \cdot \vec{x}\right) \times \vec{\pi} \right) \times \vec{\pi} \right]
+ i \, \sigma^k \, \epsilon^i{}_{jk} \left[ \nabla_i
\left(\vec{\nabla}\left(\vec{a} \cdot \vec{x}\right) \times \vec{\pi} \right)^j\right]
\right] \right.
\nn
&&{} + {2 \hbar \over |\vec{\pi}|^2} \left[\left(\nabla_k R'^k \right) + i \,
\sigma^k \, \epsilon^i{}_{jk} \left(\nabla_i R'^j \right) + {2 \over \hbar} \,
\vec{\sigma} \cdot \left(\vec{R'} \times \vec{\pi}\right) \right] \vec{\nabla} \Phi_{\rm G} \cdot \vec{\pi}
- 2 \, \vec{\sigma} \cdot \vec{\nabla} \left(\vec{\nabla}\Phi_{\rm G} \cdot \vec{\pi}\right)
\nn
&&{} + \left. {2 \hbar \over |\vec{\pi}|^2} \left[\vec{R'} \cdot \vec{\nabla}
\left(\vec{\nabla}\Phi_{\rm G} \cdot \vec{\pi}\right) + i \,
\vec{\sigma} \cdot \left[\vec{R'} \times \vec{\nabla}\left(\vec{\nabla}\Phi_{\rm G} \cdot \vec{\pi}\right)
\right] \right] \right]\beta.
\label{hdot-CT1=}
\ee
After neglecting the non-Hermitian term
$- \left(i \hbar/2\right) (\vec{\alpha} \cdot \vec{a})$ in (\ref{H-accel=}),
it is straightforward to confirm that (\ref{hdot-CT0=}) and (\ref{hdot-CT1=}) reduce to (\ref{hdot=}) in the
limit as $q \ \rightarrow \ 0$.


\section{Spin Evolution}
\setcounter{equation}{0}

It is useful to evaluate the spin evolution of a spin-1/2 particle
in non-inertial motion, such as that due to a circular orbit in an
ideal storage ring. This requires that the operator expressions
for the helicity transition rate must be converted into
amplitudes and projected into the laboratory frame.
Suppose that a beam of spin-1/2 particles follows a circular orbit
in an idealized storage ring, allowing for vertical and horizontal
fluctuations about the beam's mean trajectory. Then, by adopting
cylindrical co-ordinates $(r, \theta, z, \tau)$ to describe an
accelerated frame tangent to the beam orbit, where
\be
P^1 & = & -i \hbar \, {\partial \over \partial r}, \qquad
P^2 \ = \ {-i \hbar \over r} \, {\partial \over \partial \theta}, \qquad
P^3 \ = \ -i \hbar \, {\partial \over \partial z},
\label{p-cylind=}
\ee
it follows that $\vec{R} = \left(0, 0, -{1 \over r} \, P^2\right)$.
Assuming a mean orbital radius of $r_0$ and orbital frequency of $\omega_0$ in the
laboratory frame,
the relationship between the accelerated frame and the laboratory frame in Cartesian
co-ordinates $(x, y, z, t)$,
whose origin is at the centre of the storage ring, is given by \cite{Bell,Mashhoon}
\be
x & = & \left(r_0 + \delta r \right) \cos \left(\gamma \omega_0 \tau \right)
- \gamma \left(r_0 \, \delta \theta \right)  \sin \left(\gamma \omega_0 \tau \right)
\nn
y & = & \left(r_0 + \delta r \right) \sin \left(\gamma \omega_0 \tau \right)
+ \gamma \left(r_0 \, \delta \theta \right)  \cos \left(\gamma \omega_0 \tau \right)
\nn
t & = & \gamma \left(\tau + r_0^2 \, \omega_0 \, \delta \theta \right),
\label{LT=}
\ee
where $\delta r$ and $\delta \theta$ are the radial and angular fluctuations about the
mean orbit,
$\gamma = \left(1 - \omega_0^2 \, r_0^2\right)^{-1/2}$, and $\tau$ is the proper time.

To obtain the spin-flip transition amplitude, it is necessary to adopt the
Dirac representation and evaluate it in the instantaneous rest frame of the particle.
Therefore, given the corresponding ket vectors \cite{Sakurai}
\be
\left| + \right\rangle_{\rm up} & \equiv &
\left(\begin{array}{c}
\left(\begin{array}{c}
1  \\  0 \end{array} \right) \\ \\
\left(\begin{array}{c}
0  \\  0 \end{array} \right)
\end{array} \right) 
\qquad
\left| - \right\rangle_{\rm up} \ \equiv \
\left(\begin{array}{c}
\left(\begin{array}{c}
0  \\  1 \end{array} \right) \\ \\
\left(\begin{array}{c}
0  \\  0 \end{array} \right)
\end{array} \right) 
\nn
\left| + \right\rangle_{\rm dn} & \equiv &
\left(\begin{array}{c}
\left(\begin{array}{c}
0  \\  0 \end{array} \right) \\ \\
\left(\begin{array}{c}
1  \\  0 \end{array} \right)
\end{array} \right) 
\qquad
\left| - \right\rangle_{\rm dn} \ \equiv \
\left(\begin{array}{c}
\left(\begin{array}{c}
0  \\  0 \end{array} \right) \\ \\
\left(\begin{array}{c}
0  \\  1 \end{array} \right)
\end{array} \right), 
\label{ket=}
\ee
it is shown that the spin-flip transition amplitude due to some quantum operator $Q$ is
\be
\langle Q \rangle & \equiv & \langle \mp | Q | \pm \rangle_{\rm up/dn}.
\label{<Q>=}
\ee
Then for some arbitrary spin-independent quantum number $\vec{K}$ coupled to
$\vec{\sigma}$, it follows that
\be
\left\langle (\vec{\sigma} \cdot \vec{K}) \right\rangle & = &
\left(\hat{\vec{x}}^1 \pm i \, \hat{\vec{x}}^2 \right) \cdot \vec{K}
\label{amplitude1=}
\nl
\left\langle (\vec{\sigma} \cdot \vec{K})\beta \right\rangle & = &
\pm \left(\hat{\vec{x}}^1 \pm i \, \hat{\vec{x}}^2 \right) \cdot \vec{K},
\label{amplitude2=}
\ee
where all other matrix elements vanish, and the overall $\pm$ in (\ref{amplitude2=}) denotes
the sign of the contribution specific to the up/dn state, respectively.

It is straightforward to verify \cite{Mashhoon2} that the unit vectors
$\hat{\vec{x}}^1(\tau)$ and
$\hat{\vec{x}}^2(\tau)$ are related to the time-independent Cartesian unit vectors
$\hat{\vec{x}}$ and $\hat{\vec{y}}$ in the laboratory frame by
\be
\hat{\vec{x}}^1(\tau) \pm i \, \hat{\vec{x}}^2(\tau) & = &
\left(\hat{\vec{x}} \pm i \, \hat{\vec{y}} \right) e^{\mp i \gamma \omega_0 \tau}
\ = \ \left(\hat{\vec{x}} \pm i \, \hat{\vec{y}} \right) e^{\mp i \omega_0 t}.
\label{unitvectors=}
\ee
With (\ref{amplitude1=}) and (\ref{amplitude2=}), it is formally shown that the
rate of change of the spin-flip transition is
\be
\left\langle {d h(t) \over d t} \right\rangle & = &
{1 \over |\vec{\pi}|}
\left(\hat{\vec{x}}^1(t) \pm i \, \hat{\vec{x}}^2(t)\right) \cdot
\left[\vec{\Lambda}^0 \pm \vec{\Lambda}^1 \right] \ = \
{1 \over |\vec{\pi}|}
\left(\hat{\vec{x}} \pm i \, \hat{\vec{y}}\right) \cdot
\left[\vec{\Lambda}^0 \pm \vec{\Lambda}^1 \right] e^{\mp i \omega_0 t},
\label{<hdot>=}
\ee
where $\vec{\Lambda}^0$ and $\vec{\Lambda}^1$ are the amplitudes corresponding to
(\ref{amplitude1=}) and (\ref{amplitude2=}), respectively.
It then follows that the spin evolution is
\be
\langle h (t) \rangle  & = &
{1 \over |\vec{\pi}|} \left(\hat{\vec{x}} \pm i \, \hat{\vec{y}} \right)
\cdot \vec{\pi} + \int_0^t \left\langle {d h (t') \over d t} \right\rangle  dt'
\nn
& = & {1 \over |\vec{\pi}|} \left(\hat{\vec{x}} \pm i \, \hat{\vec{y}} \right) \cdot
\left[\vec{\pi} \pm {i \over \omega_0} \left( \vec{\Lambda}^0 \pm \vec{\Lambda}^1 \right)
\left(e^{\mp i \omega_0 t} - 1\right) \right].
\label{h-evolution=}
\ee

By evaluating (\ref{amplitude1=}) and (\ref{amplitude2=}) for each of the cases under
consideration,
the amplitudes corresponding to the Dirac Hamiltonian in its original representation,
and also its FW- and CT-transformed counterparts, can be obtained.
Therefore, from (\ref{hdot=}), it is shown that
\be
\vec{\Lambda}^0_{\rm Dirac} & = & \vec{\omega} \times \vec{\pi}
+ {i \hbar \over 2} \, \epsilon^i{}_{jk} \left( \nabla_i \omega^j \right) \hat{\vec{x}}^k
+ \left(\vec{\omega} \times \vec{x} \right) \times \vec{R}
+ \epsilon_{klm} \left[\vec{\nabla} \left(\omega^l \, x^m\right)\right] \pi^k
\nn
&  &{} - e \left(\vec{\omega} \times \vec{x} \right)^k
\left[\left(\nabla_k A_l\right) - \left(\nabla_l A_k\right) \right] \hat{\vec{x}}^l
- e \, \vec{\nabla} \varphi - \vec{\nabla}(\nabla_0 \Phi_{\rm G})
\label{amplitude1-Dirac=}
\nl
\vec{\Lambda}^1_{\rm Dirac} & = & - m \, \vec{\nabla} \left(\vec{a} \cdot \vec{x}\right) +
\vec{\Lambda}^1_\kappa
\label{amplitude2-Dirac=}
\nl
\vec{\Lambda}^1_\kappa & = & {\kappa e \hbar \over 2m} \left[{2 \over \hbar} \left(1 + \vec{a} \cdot \vec{x} \right)
\vec{B} \times \vec{\pi} + i \left(1 + \vec{a} \cdot \vec{x} \right) \epsilon^i{}_{jk}
\left( \nabla_i  B^j \right) \hat{\vec{x}}^k
+ i \, \vec{\nabla} \left(\vec{a} \cdot \vec{x}\right) \times \vec{B} \right],
\label{amplitude3-Dirac=}
\ee
for the helicity evolution involving the original Dirac
Hamiltonian, where $\vec{\omega} = \gamma^2 \, \omega_0 \,
\hat{\vec{z}}$ \cite{Bell,Irvine} in the rotating frame of
reference. In an ideal storage ring, the corrections proportional
to $\vec{a}$ found in (\ref{amplitude3-Dirac=}) and several other
expressions for $\Lambda $ given below are really of second
order because $B = m \omega /e$. The latter relationship also
ensures that $\Lambda_{\kappa}^{1}$ is mass independent. The
remaining terms in (\ref{amplitude3-Dirac=}) do not contribute to
$g-2$ experiments because of geometrical constraints
or because $B$ is uniform. They would
contribute, however, to experiments where these constraints were
relaxed.

The first and second terms in (\ref{amplitude1-Dirac=}) originate
from the Mashhoon coupling in the Hamiltonian and the second term
would not occur in experiments with $\omega$ constant. The first
and fourth term play an important role in $g-2$ experiments and
are discussed below.
The sixth term is well known and can also be found
in the calculation of Sakurai \cite{Sakurai}.
Together, the fifth and sixth terms form something akin to the
Lorentz force in a rotating frame.

The third term
vanishes in Cartesian coordinates because that case involves only
linear momenta and the commutator in (\ref{Rvec=}) therefore
vanishes. However, when the coordinates are not Cartesian, the
commutator of the momenta mixes linear momentum and angular
momentum components and, in general, does not vanish. Its
contribution to the precession equation is $ \vec{\sigma}\cdot
\left[\left(\vec{\omega}\times \vec{x}\right)\times
\vec{R}\right]$ and has the same dimensions as the term $
\vec{\sigma}\cdot \left(\vec{\omega}\times \vec{p}\right)$ that
appears in the Thomas-BMT equation \cite{Montague}. The third term
may be small for the geometry of $g-2$ experiments, but not so for
other types of spin motion like those considered in spin rotators
and Siberian snakes.

The last term in (\ref{amplitude1-Dirac=}) is new and has interesting consequences.
In fact, it follows from (\ref{<hdot>=})-(\ref{amplitude3-Dirac=}) that in the absence of
electromagnetic potentials and in the limit as $m \ \rightarrow \ 0,$
\be
\left\langle {d h \over d t} \right\rangle_{m = 0}
& = & - {1
\over |\vec{\pi}|} \, \left(\hat{\vec{x}}^1 \pm i \,
\hat{\vec{x}}^2 \right) \cdot \vec{\nabla}(\nabla_0 \Phi_{\rm G}),
\label{hdot(m=0)=}
\ee
where, for the case of a Cartesian co-ordinate frame,
\be
\nabla_{i} \left(\nabla_{0}\Phi_{\rm G} \right)
& = & -{1 \over 2} \left[\gamma_{00,i} \, p^0 +
\left(\gamma_{0j,i} + \gamma_{ij,0} - \gamma_{0i,j}\right)p^j \right].
\label{nabla}
\ee
For instance, given a stationary metric of the type shown by Hehl
and Ni \cite{Hehl}, we obtain
\be
\nabla_{i} \left(\nabla_{0} \Phi_{\rm G}\right)
& = & -\left(\vec{a}\cdot\vec{x}\right)_{,i} \, p^0 -
{1 \over 2} \left[\epsilon_{jkl} \left(\omega^k \, x^l\right)_{,i} -
\epsilon_{ikl} \, \left(\omega^k \, x^l\right)_{,j} \right]p^j
\nn
& \approx & a_i \, p^0 + \epsilon_{ijk} \, \omega^j \, p^k \ \neq \ 0.
\label{nabla1}
\ee
This result is somewhat surprising, but not unlike the electromagnetic case where
the presence of an electric field also violates helicity conservation \cite{Sakurai}.
The fact that $\dot{h} \neq 0$ here challenges commonly held views that the
helicity is a constant of motion for massless particles \cite{Itzykson}.
However, particles are known to acquire an effective mass when
acted upon by inertia-gravitation.  By multiplying (\ref{covDirac=})
on the left by $ \left(-i
\gamma^{\nu}\left(x\right)D_{\nu}-m/\hbar\right)$ and using the
relations $ \left[D_{\mu},D_{\nu}\right]=-i
\sigma^{\alpha\beta}R_{\alpha\beta\mu\nu},
\sigma^{\mu\nu}\sigma^{ab}R_{\mu\nu\alpha\beta}=2 R$, where
$R$ is the Ricci curvature scalar, we can obtain
\be
\left(g^{\mu\nu}D_{\mu}D_{\nu} - \frac{R}{4} +
\frac{m^{2}}{\hbar^{2}}\right)\psi\left(x\right) & = & 0.
\label{SOD}
\ee
Weyl \cite{weyl} was the first one to suggest that $m_{\rm eff} \equiv
\hbar \sqrt{\left(m/\hbar\right)^{2} - R/4}$ behaves as an
effective mass. Notice that $m_{\rm eff} \neq 0$ even when $m = 0$, and
that $R \neq 0 $ when it represents, in its linearized form, pure
inertia.

A second interesting result follows from the same equations.
By choosing Cartesian co-ordinates so that the term
$(\vec{\omega}\times \vec{x})\times \vec{R}$ vanishes, dropping
second order terms in $\omega$, and combining the remaining first
and fourth terms in (\ref{amplitude1-Dirac=}) with the term
$\left(\kappa e/m \right)\vec{B}\times \vec{\pi}$ in
(\ref{amplitude2-Dirac=}), we obtain
\be
\left\langle\frac{d h}{d t}\right\rangle
& \simeq &
\frac{1}{|\vec{\pi}|}\left(\hat{\vec{x}}^{1} \pm i \hat{\vec{x}}^{2}\right) \cdot
\left[\pm \, \frac{\kappa e}{m} \, \vec{B} \times \vec{\pi} +
\epsilon_{ijk} \left(\vec{\nabla}\omega^i\right)x^j \, \pi^k \right].
\label{inhom}
\ee
The Mashhoon term therefore disappears irrespective of whether $
\omega $ is constant or not. The first term proportional to
$\kappa $ and with $ B = m \omega/e $ on the right hand side of
(\ref{inhom}) is the term normally measured in $ g-2 $
experiments. If $ \omega $ is inhomogeneous, then the second
term
also contributes to the helicity precession. This term can then be
neglected only for particular geometrical configurations of the
parameters involved.

From (\ref{hdot-FW0=}) and (\ref{hdot-FW1=}), the corresponding expressions for the
FW-transformed Hamiltonian are
\be
\vec{\Lambda}^0_{\rm FW} & = & \vec{\Lambda}^0_{\rm Dirac} +
{1 \over 2m^2} \, \epsilon_{ijk} \left[\left(\vec{\omega} \times \vec{\pi} \right) \times \vec{\pi}
+ e \left(2\kappa \, \vec{E} - \vec{\nabla} \varphi\right)\times \vec{\pi} -
\vec{\nabla}(\nabla_0 \Phi_{\rm G}) \times \vec{\pi}\right]^i \, \pi^j \, \hat{\vec{x}}^k
\nn
&  &{} + {\hbar^2 \over 8m^2} \left[ \vec{\nabla}
\left[\vec{\nabla} \cdot \vec{\nabla} \left[\left(\vec{\omega} \times \vec{x} \right)
\cdot \vec{\pi} \right] + \left[\nabla_k \left(\vec{\omega} \times \vec{\pi} \right)^k\right]
+ e \left[ \nabla_k  \left(2 \kappa \, E^k + \nabla^k \varphi \right)\right] -
\vec{\nabla} \cdot \vec{\nabla}(\nabla_0 \Phi_{\rm G}) \right] \right.
\nn
&  &{} - \left. \left[ \left(\vec{\omega} \times \vec{x} \right)^j \left[\nabla_j \nabla_k
\left(\nabla_i \pi^i \right) \right]\hat{\vec{x}}^k - \vec{\omega} \times \left[\vec{\nabla}
\left(\nabla_k \pi^k \right)\right] \right] \right]
\nn
&  &{} + {\hbar^2 \over 4m^2} \left[ \left(\vec{\omega} \times \vec{x} \right) \cdot \vec{\nabla}
\left[\left(\nabla_1 R^3\right)\hat{\vec{x}}^2 - \left(\nabla_2 R^3\right)\hat{\vec{x}}^1 -
 e \left(\left(\nabla^j \nabla_j A_k\right) - \left(\nabla^j \nabla_k A_j \right)\right)\hat{\vec{x}}^k
 \right] \right.
\nn
&  &{} + \left[R^3 \left[ \epsilon^2{}_{jk}
\left(\nabla_1 \omega^j\right) - \epsilon^1{}_{jk} \left(\nabla_2 \omega^j\right)\right]
+ \epsilon_{ijk} \,
e \left(\left(\nabla_m A^i\right) - \left(\nabla^i A_m\right) \right)
\left(\nabla^m \omega^j\right)\right]\hat{\vec{x}}^k
\nn
&  &{} + \left. \omega^i \left[\epsilon^2{}_{ik}
\left(\nabla_1 R^3\right) - \epsilon^1{}_{ik} \left(\nabla_2 R^3\right)
+ \epsilon_{ijk} \,
e \left(\left(\nabla^m \nabla_m A^j\right) - \left(\nabla^m \nabla^j A_m\right)\right)
\right]\hat{\vec{x}}^k \right]
\nn
&  &{} + {i \hbar \over 4m^2} \, \epsilon^i{}_{jk} \, \nabla_i
\left[\left(\vec{\omega} \times \vec{\pi} \right) \times \vec{\pi}
+ e \left(2\kappa \, \vec{E} - \vec{\nabla} \varphi\right)\times \vec{\pi} -
\left[\vec{\nabla}(\nabla_0 \Phi_{\rm G})\right] \times \vec{\pi} \right]^j \hat{\vec{x}}^k
\nn
&  &{} - {i \hbar^2 \over 2m^2}\left[
\left(\vec{\omega} \times \vec{x} \right) \cdot \vec{\nabla}
\left[R^3 \left(\pi^1 \, \hat{\vec{x}}^2 - \pi^2 \, \hat{\vec{x}}^1 \right) +
e \left[\left(\nabla_j A_k\right) - \left(\nabla_k A_j \right)\right] \pi^j \, \hat{\vec{x}}^k \right] \right.
\nn
&  &{} + \left. \omega^i \left[R^3 \left(\epsilon^2{}_{ik} \, \pi^1 - \epsilon^1{}_{ik} \, \pi^2 \right)
+ \epsilon_{ijk} \, e \, \left[\left(\nabla_m A^j\right) - \left(\nabla^j A_m\right)
\right] \pi^m \right]\hat{\vec{x}}^k\right]
\label{amplitudes-FW0=}
\nl
\vec{\Lambda}^1_{\rm FW} & = & -\left(1 + \vec{a} \cdot \vec{x} \right)
\left[{1 \over 2m} \, \vec{\nabla} \left(\vec{\pi} \cdot \vec{\pi}\right)
+ {1 \over m} \, \vec{\sigma} \cdot \left(\vec{R'} \times \vec{\pi} \right)
+ {i \hbar \over 2m} \, \epsilon^i{}_{jk} \left(\nabla_i R'^j \right) \hat{\vec{x}}^k \right]
\nn
&  &{} - \vec{\nabla}\left(\vec{a} \cdot \vec{x}\right) \left[m + {1 \over 2m} \, \vec{\pi} \cdot \vec{\pi} \right]
+ {\kappa e \over m} \, \vec{\sigma} \cdot \left(\vec{B} \times \vec{\pi}\right)
+ {i \kappa e \hbar \over 2m} \, \epsilon^i{}_{jk} \left( \nabla_i B^j \right)
\hat{\vec{x}}^k  - {\hbar \over 2m} \, \epsilon^{ij}{}_k
\left( \nabla_i \, \nabla_j \Phi_{\rm G} \right) \hat{\vec{x}}^k
\nn
&  &{} - {i \hbar \over 2m} \left[ \vec{\nabla}\left(\vec{a} \cdot \vec{x}\right) \times \vec{R'}
+ \left(\nabla^j \left[\nabla_j \nabla_k - \nabla_k \nabla_j \right]\Phi_{\rm G} \right) \hat{\vec{x}}^k \right]
\nn
&  &{} - {1 \over 2m}
\left[ \left(\vec{\nabla}\left(\vec{a} \cdot \vec{x}\right)
\times \vec{\pi} \right)\times \vec{\pi} + 2 \, \vec{\nabla}(\vec{\nabla} \Phi_{\rm G} \cdot \vec{\pi}) \right]
- {1 \over m} \left(\left[\nabla_j \nabla_k - \nabla_k \nabla_j \right]\Phi_{\rm G} \right) \pi^j \, \hat{\vec{x}}^k
\nn
&  &{} - {\hbar^2 \over 8m} \,
\vec{\nabla} \left(\vec{\nabla} \cdot \vec{\nabla}\left(\vec{a} \cdot \vec{x}\right) \right) -
{i \hbar \over 4m} \, \epsilon^i{}_{jk} \left[
\nabla_i \left(\vec{\nabla}\left(\vec{a} \cdot \vec{x}\right)
\times \vec{\pi} \right)^j \right] \hat{\vec{x}}^k.
\label{amplitudes-FW1=}
\ee

Again assuming a local Cartesian frame,
the leading-order contributions to the amplitude in the low-energy approximation, of
order $1/m$,
come from the second and sixth terms of (\ref{amplitudes-FW1=}), which yield the total
magnetic moment term $\left[\left(1 + \kappa\right) e/m \right] \vec{\sigma} \cdot
\left(\vec{B} \times \vec{\pi} \right)$.
Other noteworthy leading-order contributions are due to the eleventh term in
(\ref{amplitudes-FW1=}), which come from the acceleration-induced spin-orbit coupling term
first found by Hehl and Ni \cite{Hehl,Singh1}, and also the twelfth and thirteenth terms
due to the gravitational energy redshift term found by Singh and Papini
\cite{Singh1}.
As for terms of order $1/m^2$, the leading contributions \cite{Singh1} are due to the second
to fourth terms found in (\ref{amplitudes-FW0=}), namely the spin-orbit coupling from
the Mashhoon effect, electric field, and gravitational field, respectively.
In addition, the seventh and eighth terms of (\ref{amplitudes-FW0=}) identify contributions
from the Darwin energy terms due to electromagnetism \cite{Sakurai} and gravitation, also
first found by Singh and Papini \cite{Singh1}, and later by Obukhov \cite{Obukhov}.
All other terms in (\ref{amplitudes-FW0=}) and (\ref{amplitudes-FW1=}) involving gradients
of primarily small quantities can be safely regarded as negligible by comparison.

From (\ref{hdot-CT0=}) and (\ref{hdot-CT1=}), it is shown that the amplitudes for the
CT-transformed Hamiltonian are
\be \vec{\Lambda}^0_{\rm CT} & = &
\vec{\Lambda}^0_{\rm Dirac} + {1 \over |\vec{\pi}|} \, {\kappa e \hbar \over 2m}  \left[
\vec{\nabla} \left(\vec{a} \cdot \vec{x} \right) \vec{B} \cdot \vec{\pi}
- i \, \vec{\nabla} \left(\vec{a} \cdot \vec{x} \right)
\times \left(\vec{E} \times \vec{\pi} \right) \right.
\nn
&  &{} + \left. \left(1 + \vec{a} \cdot \vec{x} \right) \left[\vec{\nabla}
\left(\vec{B} \cdot \vec{\pi}\right) - {2 \over \hbar}
\left(\vec{E} \times \vec{\pi} \right) \times \vec{\pi}
- i \, \epsilon^i{}_{jk} \left[ \nabla_i \left(\vec{E} \times \vec{\pi}
\right)^j \right] \hat{\vec{x}}^k \right] \right]
\label{amplitudes-CT0=}
\nl
\vec{\Lambda}^1_{\rm CT} & = & \vec{\Lambda}^1_\kappa + {\hbar
\over 2|\vec{\pi}|^2} \sqrt{|\vec{\pi}|^2 + m^2}
\left[ {2 \over \hbar}\left(\vec{\nabla}\left(\vec{a} \cdot \vec{x}\right) \times
\vec{\pi} \right) \times \vec{\pi}
+ i \, \epsilon^i{}_{jk} \left[
\nabla_i \left(\vec{\nabla}\left(\vec{a} \cdot \vec{x}\right)
\times \vec{\pi} \right)^j \right] \hat{\vec{x}}^k \right]
\nn
&  &{} +
{\hbar \over |\vec{\pi}|^3} \left[{2 \over \hbar} \,
\left(\vec{R'} \times \vec{\pi}\right) + i \, \epsilon^i{}_{jk}
\left(\nabla_i R'^j \right) \, \hat{\vec{x}}^k
\right]\vec{\nabla}\Phi_{\rm G} \cdot \vec{\pi}
\nn
&  &{}
- {1 \over |\vec{\pi}|} \, \vec{\nabla}\left(\vec{\nabla}\Phi_{\rm G}
\cdot \vec{\pi}\right) + {i \hbar \over |\vec{\pi}|^3} \, \vec{R'}
\times \vec{\nabla}\left(\vec{\nabla}\Phi_{\rm G} \cdot
\vec{\pi}\right).
\label{amplitudes-CT1=}
\ee
Given that the CT-Hamiltonian is an ultrarelativistic approximation of the original Dirac
Hamiltonian, we expect that many of the terms in (\ref{amplitudes-CT0=}) and
(\ref{amplitudes-CT1=}) will be small compared to the contributions due to the original
Dirac Hamiltonian if we disregard those due to inhomogeneous fields.
Nonetheless, there are a few noteworthy terms which should make a meaningful contribution.
One of them is the fifth term in (\ref{amplitudes-CT0=}), a spin-orbit coupling due to the
electric field which yields a term ${1 \over |\vec{\pi}|} \left(\kappa e/m\right)
\left(\vec{E} \times \vec{\pi}\right) \times \vec{\pi}$.
The other contributions of note are the second, fourth, and sixth terms of
(\ref{amplitudes-CT1=}), which are due to the ultrarelativistic analogues of the
Hehl-Ni spin-orbit coupling and gravitational energy redshift terms found in the
low-energy approximation.

\section{Chiral Transition Rate for a Relativistic Spin-1/2 Particle in
a Gravitational Field}

\setcounter{equation}{0}

It is of interest to also study the effect of inertia on the
chirality precession of spin-1/2 particles, for comparison with
the helicity precession. Given that the $\gamma^5$ operator in the
chiral representation is
\be
\gamma^5 \ = \ \left(
\begin{array}{cc}
1 & 0 \\ 0 & -1
\end{array} \right),
\label{5.1}
\ee
the projection operators for isolating right- and left-handed states
are defined as
\be
P_{\rm R} & \equiv & {1 \over 2} \, (1 + \gamma^5),
\qquad
P_{\rm L} \ \equiv \ {1 \over 2} \, (1 - \gamma^5),
\label{5.3}
\ee
%
%
\be
\psi & \equiv & \left(
\begin{array}{c}
\varphi_{\rm R} \\ \varphi_{\rm L}
\end{array} \right).
\label{5.4}
\ee
Equivalently, (\ref{5.3}) can be written as
\be
P_{\pm} & = & {1 \over 2} \, (1 \pm \gamma^5).
\label{proj=}
\ee

Applying the CT transformation on (\ref{proj=}) leads to
\be
P_{\rm CT \, \pm} & \approx & P_{\pm} \pm {\omega(q) \over 2 |\vec{\pi}|} \, \gamma^5 \,
\beta \left(\vec{\alpha} \cdot \vec{\pi}\right).
\label{P-CT=}
\ee
Then, to leading order in $q$,
\be
\dot{P}_{\rm CT \, \pm} & \approx &  \pm \, q \left[
{i \over \hbar} \, H_1 \, \gamma^5 \, \beta - {1 \over 2}
\left(1 - {\hbar \over |\vec{\pi}|^2} \, \vec{\sigma} \cdot \vec{R'} \right)
\left\{ {i \over \hbar} \left[H_0, h \right] +
{1 \over |\vec{\pi}|} \left(\vec{\alpha} \cdot \vec{\pi} \right) \cdot
{i \over \hbar} \left[H_0, \beta \right] \beta \, \gamma^5 \right\}
\right] \beta.
\label{P-CT-dot=}
\ee
Explicitly, the chirality transition rate (\ref{P-CT-dot=}) is given by (\ref{hdot-CT0=}) and
\be
\lefteqn{
{1 \over |\vec{\pi}|} \left(\vec{\alpha} \cdot \vec{\pi} \right) \cdot
{i \over \hbar} \left[H_0, \beta \right] \beta \ = \ {2 \over |\vec{\pi}|^2} \left[
\sqrt{|\vec{\pi}|^2 + m^2} \left[\vec{\nabla}\left(\vec{a} \cdot \vec{x}\right) \cdot \vec{\pi}
+ i \,\vec{\sigma}  \cdot \left[\vec{\nabla}\left(\vec{a} \cdot \vec{x}\right) \times \vec{\pi} \right] \right]
\right. }
\nn
&&{} - \left. {q^3 \over \sqrt{1 + q^2}} \, {\hbar \over 2m}
\left[\left[\vec{\sigma} \cdot \vec{\nabla}\left(\vec{a} \cdot \vec{x}\right)\right]
\vec{R'} \cdot \vec{\pi}
+ i \, \vec{\nabla}\left(\vec{a} \cdot \vec{x}\right) \cdot \left(\vec{R'} \times \vec{\pi} \right)
- \vec{\sigma} \cdot \left[\vec{\nabla}\left(\vec{a} \cdot \vec{x}\right) \times
\left(\vec{R'} \times \vec{\pi} \right) \right] \right] \right]
\nn
&&{} + {2i \over \hbar |\vec{\pi}|^2} \left(1 + \vec{a} \cdot \vec{x} \right)
\left[\sqrt{|\vec{\pi}|^2 + m^2} \left(\vec{\pi} \cdot \vec{\pi} + \hbar \, \vec{\sigma} \cdot \vec{R'} \right)
- {q^3 \over \sqrt{1 + q^2}} \, {\hbar \over 2m} \left[ \vec{\sigma} \cdot \left[
-i \hbar \, \vec{\nabla} \left(\vec{R'} \cdot \vec{\pi}\right) +
\left(\vec{R'} \cdot \vec{\pi}\right) \vec{\pi} \right] \right. \right.
\nn
&&{} + \left. \left. \hbar \left[\left[\nabla_k \left(\vec{R'} \times \vec{\pi}\right)^k \right]
+ i \, \sigma^k \, \epsilon^i{}_{jk} \left[ \nabla_i \left(\vec{R'} \times \vec{\pi}\right)^j \right] \right] +
\vec{\sigma} \cdot \left[\left(\vec{R'} \times \vec{\pi}\right) \times \vec{\pi} \right] \right] \right]
\nn
&&{} + {2 \over |\vec{\pi}|} \left[
\left(\vec{\nabla} \cdot \vec{\nabla} \Phi_{\rm G}\right) +
i \, \sigma^k \, \epsilon^{ij}{}_k \left( \nabla_i \, \nabla_j \Phi_{\rm G}\right) \right] +
{2i \over \hbar |\vec{\pi}|} \left[\vec{\nabla}\Phi_{\rm G} \cdot \vec{\pi} - i \,
\vec{\sigma} \cdot \left[\vec{\nabla}\Phi_{\rm G} \times \vec{\pi} \right] \right].
\label{proj+=}
\ee

Excluding the contributions from the anomalous magnetic moment terms, it is clear from
(\ref{P-CT-dot=}) that
\be
\dot{P}_{\rm CT \, \pm}|_{m = 0} & = &  0,
\label{proj(m=0)=}
\ee
and so the chirality is a constant of the motion for massless
particles. This difference between (\ref{hdot(m=0)=}) and
(\ref{proj(m=0)=}) strongly suggests that helicity and chirality
describe entirely different physical processes, and their
respective interpretations may require closer investigation. A
comprehensive study of chirality transitions in a Schwarzschild
field is presented in \cite{Singh2}.

\section{Conclusions}
Inertial-gravitational fields affect quantum particles in
different ways. They interact with particle spins and give rise to
quantum phases that can be measured in principle by
interferometric means. In this case $ \Phi_{\rm G}$ must be
calculated over a closed space-time path that can be obtained, for
instance, by comparing the phase of a particle at the final
position $P_{\it f}$ at the final time $ t_{\it f}$ with that of
an identical particle at the the same final point $P_{\it f}$, but
at the initial time $t_{\it i}$.

Through the Hamiltonian, the fields can also affect the energy
levels and the time evolution of observables. In the latter case,
inertial fields change the helicity and chirality of particles in
ideal storage rings. This has been studied in some detail in
Sections 3 to 5. The results independently confirm that the
spin-rotation coupling compensates the much larger contribution
that comes from the $g=2$ part of the magnetic moment of a pure
Dirac particle. Without this cancellation, $g-2$ experiments may
be more difficult to perform with the present accuracy of 0.7 {\it
ppm} \cite{BNL2}.

In the more general case of an inhomogeneous $\vec{\omega}$, the
Mashhoon term per se essentially disappears, but a new term $1/|\pi| \,
\epsilon^i{}_{jk}\left(\omega_{i,1}+i \, \omega_{i,2}\right)x^{j} \, \pi^{k}$
contributes to the spin precession. In the lowest approximation,
the spin precesses with the same angular frequency $\omega$ of the
particle itself. The ratio of this new term to $\omega$ is $\simeq
\left[\left(\nabla \omega\right)/\omega \right]x$.
It may be possible to
conceive of a physical situation in which this term can be
observed, which would extend our knowledge of rotational inertia.

A second interesting result is represented by (\ref{hdot(m=0)=}),
which states that helicity is not conserved in the presence of
\emph{first-order} inertial and gravitational fields, even when the
mass of the particle vanishes. The corresponding result
(\ref{proj(m=0)=}) for chirality gives a vanishing result. It
seems, therefore, that first-order inertia-gravitation can
distinguish between helicity and chirality. This may be due to the
approximation itself. It has, in fact, been mentioned that
particles acquire an effective mass when immersed in gravitational
fields. This result extends to inertia because, in the weak field
approximation, $ R = -{1 \over 2} \,
\partial_{\nu}\partial^{\nu}\gamma_{\mu}{}^{\mu}$ need not vanish.
In the general case, however, $R=0$ rigorously for inertial fields,
but not necessarily so for \emph{true} gravitational fields.
Finally, it may be exceedingly difficult to subject massless
fermions to acceleration and rotation as required, for instance,
by (\ref{nabla}) and (\ref{nabla1}). In this sense, it may be said that
massless fermions strive to conform to a sort of helicity
conservation.

\section{Acknowledgements}
This work was supported in part by the Natural Sciences and Engineering
Research Council of Canada.


\begin{thebibliography}{99}
\bibitem{DW} DeWitt B S 1966 {\it Phys. Rev. Lett.} {\bf 16} 1092
\bibitem{Papini1} Papini G 1966 {\it Nuovo Cimento} {\bf 45} 66;
1966 {\it Phys. Lett.} {\bf 23} 418; 1967 {\it Phys. Lett. A} {\bf
24} 32; 1967 {\it Nuovo Cimento B} {\bf 52} 136
\bibitem{CaiPap} Cai Y Q and Papini G 1989 {\it Class. Quantum
Grav.} {\bf 6} 407
\bibitem{Hild} Hildebrandt A F, Saffren M M 1965 {\it Proc. 9th Int. Conf.
on Low-Temp. Phys. Pt. A} 459; Hendricks J B and Rorschach H E Jr
{\it ibid.} 466; Bol M and Fairbank W M {\it ibid.} 471;
Zimmerman J E and Mercereau J E 1965 {\it Phys. Rev. Lett.} {\bf
14} 887
\bibitem{COW} Colella R, Overhauser A W and Werner S A 1975 {\it Phys.
Rev. Lett.} {\bf 34} 1472
\bibitem{GP} Papini G 2004 {\it Relativity in Rotating Frames}
ed Guido Rizzi and Matteo Luca Ruggiero (Kluwer Academic
Publishers: Dordrecht) 335
\bibitem{pap2}  Papini G 2002 {\it Phys. Rev. D} {\bf 65} 077901
\bibitem{Bell} Bell J S and Leinaas J 1987 {\it Nucl. Phys.} {\bf B284}
488
\bibitem{Cai4} Cai Y Q, Lloyd and Papini G 1993 {\it Phys. Lett.}
{\bf A178} 225
\bibitem{BNL}
Bennett G W {\it et al.} (Muon (g-2) Collaboration) 2002 {\it
Phys. Rev. Lett.} {\bf 89} 101804; Brown H N {\it et al.} (Muon
(g-2) Collaboration) 2001 {\it Phys. Rev. Lett.} {\bf 86} 2227;
2000 {\it Phys. Rev. D} {\bf 62} 091101
\bibitem{Mashhoon1} Mashhoon B 1988 {\it Phys. Rev. Lett.} {\bf 61} 2639
\bibitem{Mashhoon2} Mashhoon B 1989 {\it Phys. Lett.} {\bf A139} 103
\bibitem{Cai1} Cai Y Q and Papini G 1991 {\it Phys. Rev. Lett.} {\bf 66}
1259; 1992 {\it Phys. Rev. Lett.} {\bf 68} 3811
\bibitem{Papini2} Papini G 1994 {\it Proc. of the 5th Canadian Conference on
General Relativity and Relativistic Astrophysics} ed R B Mann and
R G McLenaghan (World Scientific: Singapore) 107
\bibitem{Mergul} Mergulh\~{a}o Carlos Jr 1994 {\it General
Relativity and Gravitation} {\bf 27} 657
\bibitem{Ald} Aldovrandi R, Matsas G E A, Novaes S F and Spehler
D 1994 {\it Phys. Rev. D} {\bf 50} 2645
\bibitem{Singh1} Singh D and Papini G 2000 {\it Nuovo Cimento} {\bf 115} 223
\bibitem{DeOliveira} De Oliveira C G and Tiomno J 1962 {\it Nuovo Cimento} {\bf 24} 672
\bibitem{Peres} Peres A 1962 {\it Suppl Nuovo Cimento} {\bf 24} 389
\bibitem{Hehl} Hehl F W and Ni W-T 1990 {\it Phys. Rev.} {\bf D42} 2045
\bibitem{Ob} Obukhov Y N 2001 {\it Phys. Rev. Lett.} {\bf 86} 192
\bibitem{Werner2} Werner S A, Staudenmann, J-L and Colella R 1979 {\it Phys.
Rev. Lett.} {\bf 42} 1103; Page L A 1975 {\it ibid.} {\bf 35} 543
\bibitem{Bonse} Bonse U and Wroblewski T 1983 {\it Phys. Rev. Lett.} {\bf 51} 1401
\bibitem{PAP} Papini G 2002 {\it Advances in the interplay between quantum and gravity physics}
                ed Peter G. Bergmann and V. de Sabbata (Kluwer
                Academic: Dordrecht) 317; {\it gr-qc/0110056}
\bibitem{Lamb} Papini G and Lambiase G  2002 {\it Phys. Lett. A} {\bf294} 175
\bibitem{Foldy} Foldy L L and Wouthuysen S A 1950 {\it Phys. Rev.} {\bf 78} 30
\bibitem{Cini} Cini M and Touschek B 1958 {\it Nuovo Cimento} {\bf 7} 422
\bibitem{Cai2} Cai Y Q and Papini G 1989 {\it Class. Quantum Grav.} {\bf
6} 407
\bibitem{Cai3} Cai Y Q and Papini G 1989 {\it Mod. Phys. Lett.} {\bf A4}
1143; 1990 {\it Class. Quantum Grav.} {\bf 7} 269; 1990 {\it Gen.
Rel. Grav.} {\bf 22} 259; Papini G 1990 {\it Quantum Mechanics in
Curved Space-Time} ed J Audretsch and V de Sabbata (Plenum Press:
New York) 473
\bibitem{Berry} Berry M V 1984 {\it Proc. of the Royal Soc., London} {\bf
A392} 45
\bibitem{DeFelice} De Felice F and Clarke C J S 1990 {\it Relativity on
Curved Manifolds} (Cambridge University Press:  New York)
\bibitem{Bjorken} Bjorken J D and Drell S D 1964 {\it Relativistic Quantum
Mechanics} (McGraw-Hill:  San Francisco)
\bibitem{Bose}
Bose S K, Gamba A and Sudarshan E C 1959 {\it Phys. Rev.} {\bf 113} 1661
\bibitem{Itzykson}
Itzykson C and Zuber J-B 1980 {\it Quantum Field Theory} (McGraw-Hill: Toronto)
\bibitem{Riley}
Riley K F, Hobson, M P and Bence S J 1997 {\it Mathematical Methods for Physics and
Engineering} (Cambridge University Press:  New York)
\bibitem{Mashhoon} Mashhoon B 2003 {\it gr-qc/0301065}
\bibitem{Irvine} Irvine W M 1964 {\it Physica} {\bf 30} 1160
\bibitem{Sakurai}
Sakurai J J 1967 {\it Advanced Quantum Mechanics} (Addison-Wesley:
New York)
\bibitem{Montague}
Montague B W 1984 {\it Phys. Rep.} {\bf 113} 1
\bibitem{weyl} Weyl H 1929 {\it Zeits. f. Physik} {\bf 56} 330
\bibitem{Obukhov}
Obukhov Y N 2002 {\it Fortsch. Phys.} {\bf 50} 711
\bibitem{Singh2} Singh D 2004 {\it gr-qc/0401044}
\bibitem{BNL2}
Bennett G W {\it et al.} (Muon (g-2) Collaboration) 2004 {\it hep-ex/0401008}

\end{thebibliography}
\end{document}